\documentclass[a4paper,10pt,twoside]{report}

\usepackage{geometry}	
\usepackage{setspace}

\everydisplay{}

\usepackage{tocloft}
\usepackage[T1]{fontenc}

\usepackage{indentfirst}

\usepackage{dashrule}

\usepackage{listings}
\usepackage[flushleft]{threeparttable}

\usepackage{fancyvrb}
\usepackage{xcolor}

\usepackage{mathtools}

\usepackage{titlesec}
\makeatletter
\def\hlinewd#1{%
	\noalign{\ifnum0=`}\fi\hrule \@height #1 %
	\futurelet\reserved@a\@xhline}
\makeatother

\usepackage{verbatim}
 
\usepackage{hyperref}
\hypersetup{colorlinks=false}
\usepackage{array}

\usepackage[latin9,utf8]{inputenc} 
\usepackage[T1]{fontenc}
\usepackage{ae}

\let\oldcdot\cdot 
\usepackage{breqn} 
\let\cdot\oldcdot 
\usepackage{etoolbox}
\patchcmd{\thebibliography}{\chapter*}{\section*}{}{}

\let\OLDthebibliography\thebibliography
\renewcommand\thebibliography[1]{
  \OLDthebibliography{#1}
  \setlength{\itemsep}{5pt}
  \setstretch{1.0}
}

\usepackage{multirow}

\usepackage{longtable}

\usepackage{scrextend}
\usepackage{ragged2e}
\usepackage{cancel}
\usepackage{float,hypcap}
\usepackage{changepage,titlesec}
\usepackage{sectsty}
\usepackage{soul}
\usepackage{bbold}
\usepackage{slashed}
\usepackage{tabularx}
\usepackage{amsmath,amsfonts,amssymb}
\usepackage{amsmath,bbm,latexsym,amssymb}
\usepackage{braket}
\usepackage{enumerate}
\usepackage{booktabs,hyperref}
\usepackage{bigints}
\usepackage{cite}

\usepackage{graphicx}
\usepackage{caption}
\usepackage{subcaption}
\usepackage{feynmp-auto}

\usepackage{color}
\definecolor{orange}{rgb}{1,0.5,0}
\definecolor{uglyblue}{RGB}{95,158,160}
\definecolor{newblue}{RGB}{128,0,0}
\definecolor{mygray}{RGB}{129,129,129}

\definecolor{myblue}{RGB}{0, 71, 171}

\graphicspath{{Figures/}}

\def\be{\begin{equation}}
\def\ee{\end{equation}}
\def\ba{\begin{alignedat}}
\def\ea{\end{alignedat}}
\def\bea{\begin{eqnarray}}
\def\eea{\end{eqnarray}}
\newcommand{\bs}{\begin{subequations}}
\newcommand{\es}{\end{subequations}}
\def\bd{\begin{dmath}} 
\def\ed{\end{dmath}} 

\def\vs{\vspace}
\def\hs{\hspace}

\definecolor{mycolor}{RGB}{165,111,29}

\def\n{\noindent}
\def\fn{\footnote}
\def\t{\text}

\def\t{\texttt}
\def\ts{\textsc}
\def\FM{\textsc{FeynMaster} \,}
\def\FR{\textsc{FeynRules} \,}
\def\FC{\textsc{FeynCalc} \,}

\def\FMM{\textsc{FeynMaster}}
\def\FRR{\textsc{FeynRules}}
\def\FCC{\textsc{FeynCalc}}

\allowdisplaybreaks

\begin{document}

\newcounter{tblEqCounter}     

\pagestyle{plain}
\hypersetup{pageanchor=false}
\hypersetup{pageanchor=true}
\pagenumbering{roman}
\setcounter{page}{1}
\pagenumbering{arabic}
\setcounter{page}{1}

\begin{flushright}
KA-TP-11-2025
\end{flushright}
\vspace*{-0.8cm}

\begin{center}
{\Large FeynMaster Manual}

\vspace{4mm}

\renewcommand*{\thefootnote}{\fnsymbol{footnote}}

Duarte Fontes\fn{duarte.fontes@kit.edu} 

\textit{Institute for Theoretical Physics,
Karlsruhe Institute of Technology,
76128 Karlsruhe, Germany}

\vspace{1mm}

Jorge C. Romão\fn{jorge.romao@tecnico.ulisboa.pt}

\renewcommand*{\thefootnote}{\arabic{footnote}}
\setcounter{footnote}{0}

\textit{Departamento de Física and CFTP, Instituto Superior Técnico\\[-2mm]
Universidade de Lisboa, Av. Rovisco Pais 1, 1049-001 Lisboa, Portugal,} 

(Dated: \today)

\vspace{1mm}
\end{center}

\vs{1.2mm}
\begin{addmargin}[12mm]{12mm}
\small
We present the manual for \textsc{FeynMaster} 2.1, a multitasking software for particle physics studies. This new version includes additional functions and is compatible with recent versions of related software.
It can be downloaded in \url{https://porthos.tecnico.ulisboa.pt/FeynMaster/}.
\end{addmargin}

\normalsize

\section{Introduction}
\label{sec:Intro}

\n \textsc{FeynMaster} \cite{Fontes:2019wqh} was introduced in 2019 as a multitasking software for particle physics studies. Combining \ts{FeynRules}~\cite{FeynRules-old,FeynRules-new}, \ts{QGRAF}~\cite{QGRAF} and \ts{FeynCalc}~\cite{FeynCalc-old,FeynCalc-new,FeynCalc-new-new,Shtabovenko:2023idz}, \FM is able to perform the totality of the following list of tasks:
\vs{1.0mm}
\begin{center}
\quad a) generation and drawing of Feynman rules; \qquad b) generation and drawing of Feynman diagrams;\\
c) generation of amplitudes; \hs{3mm} d) loop calculations; \hs{3mm} e) algebraic calculations; \hs{3mm} f) renormalization.
\end{center}
\vs{1.0mm}
It was presented not as competing with other existing software that perform some of the elements of the above list (e.g., refs. \cite{Belanger:2003sd,Cullen:2011ac,Cullen:2014yla,Lorca:2004fg,Degrande:2014vpa,FeynCalc-new,FeynCalc-old,FeynRules-new,FeynCalc-new-new,FeynRules-old,Hahn:1998yk,Hahn:2000kx,Kublbeck:1990xc,Pukhov:1999gg,QGRAF,Semenov:1996es,Semenov:1998eb,Tentyukov:1999is,Wang:2004du,Alwall:2014hca}), but rather as an alternative --- with four main advantages.
First, \ts{FeynMaster} has a hybrid character concerning automatization: not only does it automatically generate the results, but it also allows the user to act upon them through \t{\ts{Mathematica}} notebooks. Second, the complete set of analytical expressions for the counterterms in the modified minimal subtraction ($\overline{\text{MS}}$) scheme can be automatically calculated. Third, \ts{FeynMaster} includes a thorough interaction with numerical calculations through \t{\ts{Fortran}}, using \ts{LoopTools}~\cite{Hahn:1998yk} in one-loop calculations. Finally, all the printable outputs of \ts{FeynMaster} --- the complete set of Feynman rules (tree-level and counterterms), the Feynman diagrams, as well as a list containing both the expressions and computed counterterms --- are automatically written in \LaTeX \, files.

\n The general usage of \ts{FeynMaster} can be summarized in a few lines: after the user has defined the relevant directories and the model, the \ts{FeynMaster} run is controlled from a single file (\t{Control.m}). Here, the sequence of processes to study can be chosen, as well as many different options. \ts{FeynMaster} is then ready to run. The run automatically generates and opens several PDF files --- according to the options chosen in \t{Control.m} --- and creates the above-mentioned \t{\ts{Mathematica}} notebooks.

\n The code has been successfully applied to many different models \cite{Fontes:2019uld,Fontes:2019fbz,Boto:2021qgu,Jurciukonis:2021izn,Fontes:2022izp,Asteriadis:2022ras,Fontes:2022sjp,Dawson:2023oce,Boto:2024jgj}. 
\FM 2.0 \cite{Fontes:2021iue} was released in 2021 with significant improvements over the previous version. Here, we present \n \FM 2.1; this differs from the 2.0 version not only by several novel features (in particular, the possibility of including Lagrangians with up to 6-point interactions), but also by the fact that it is adapted to the most updated versions of the relevant software.

\n This note is a complete and auto-sufficient manual of \ts{FeynMaster} 2.1, and is organized as follows. In section \ref{sec:Instal}, we explain how to download and install \ts{FeynMaster}, while in the following section we describe the compatibility with other software.
Section \ref{sec:Create} is devoted to the creation of models, and section \ref{sec:Usage} to the detailed usage of \ts{FeynMaster}. Then, in section \ref{sec:Examples}, we give some examples. Finally, in section \ref{sec:Summary} we present a quick first usage of \ts{FeynMaster}.

\section{Installation}
\label{sec:Instal}

\n \ts{FeynMaster} can be downloaded in:
\begin{center}
\url{https://porthos.tecnico.ulisboa.pt/FeynMaster/}.
\end{center}
\ts{FeynRules}, \ts{QGRAF} and \ts{FeynCalc}, essential to run \ts{FeynMaster}, can be downloaded in \url{https://feynrules.irmp.ucl.ac.be/}, \url{http://cfif.ist.utl.pt/~paulo/qgraf.html} and \url{https://feyncalc.github.io/}, respectively.
After downloading \ts{FeynMaster}, the downloaded file should be extracted and the resulting folder (named \t{FeynMaster}) can be placed in any directory of the user's choice.
Then, the directories corresponding to \ts{FeynRules} and \ts{QGRAF}, as well as the one corresponding to the \ts{FeynMaster} output (i.e., the directory where the outputs of \FM shall be saved), should be defined. This must be done by writing the appropriate paths after \t{dirFR}, \t{dirQ} and \t{dirFMout}, respectively, in the beginning of the \t{RUN-FeynMaster} batch file lying inside the \t{FeynMaster} folder.%
\fn{\label{note:instal}%
These lines are commented in the batch file, and must continue to be so; besides, the beginning of these lines (with \t{dirFR}, \t{dirFR} and \t{dirFMout}) should \textit{not} be erased.
By default, the folder with the models for \ts{FeynMaster} is inside the \t{FeynMaster} folder and is named \t{Models}; this can be changed by defining a variable \t{dirFRmod} in the beginning of \t{RUN-FeynMaster} and setting a path for it. In a similar way, the folder with the models for \ts{QGRAF} (that is, the folder that shall contain the \ts{QGRAF} models, which will be automatically generated by \ts{FeynMaster}) is automatically generated and set inside the \ts{QGRAF} folder with the default name \t{Models}; as before, this can be changed by defining a variable \t{dirQmod} in the beginning of \t{RUN-FeynMaster} and setting a path for it.}

\n To test \ts{FeynMaster}, jump to section \ref{sec:Summary}, where instructions for a quick first usage are given.

\section{Software compatibility}
\label{sec:Compatibility}

\noindent The user of \ts{FeynMaster} is supposed to be familiar with both \ts{FeynRules} (in order to define new models) and \ts{FeynCalc} (in order to manipulate the final results).
There is no need to know \ts{QGRAF}, since the non-trivial part of this program --- the definition of the model --- is automatically carried through by \ts{FeynMaster}.
We verified that \ts{FeynMaster} runs properly with the latest public versions of \ts{FeynRules}, \ts{QGRAF} and \ts{FeynCalc} --- namely, versions 2.3.49, 4.0.5 and 10.1.0, respectively. It is also compatible with older versions of \ts{QGRAF} like version 3.6.6 and of \ts{FeynCalc}, like 9.3.0. For \ts{FeynRules} to work with latest \t{\ts{Mathematica}} 14, the following modification
\begin{center}
\ts{MatrixSymbol} $\to$ \ts{FRMatrixSymbol}
\end{center}
has to be made in the following three files\fn{This is necessary because the symbol \ts{MatrixSymbol} is now defined in \t{\ts{Mathematica}} 14. We thank Benjamin Fuks for this temporary solution until a new version of \ts{FeynRules} is available.}
\begin{itemize}
    \item FeynRulesPackage.m\\[-8mm]
    \item Core/MassDiagonalization.m\\[-8mm]
    \item Interfaces/AspergeInterface.m
\end{itemize}
More instructions on how to download and install QGRAF can be found in \url{https://porthos.tecnico.ulisboa.pt/CTQFT/node9.html} or in the manual that comes with the distribution \url{http://cefema-gt.tecnico.ulisboa.pt/~paulo/qgraf.html}. Finally, when using Linux or Mac \fn{We are no longer providing a version that runs on \ts{Windows}. The users interested in this implementation should contact us.}, the executable QGRAF file should be named {\t{qgraf}}.

\n To run \ts{FeynMaster}, it is also necessary to have \t{\ts{Python}}, \t{\ts{Mathematica}} and \LaTeX \, installed; links to download are \url{https://www.python.org/downloads/} \url{http://www.wolfram.com/mathematica/} and \url{https://www.latex-project.org/get/}, respectively.
We tested \ts{FeynMaster} using version 3.12 of \t{\ts{Python}} and version {14.0} of \t{\ts{Mathematica}} with the above modifications in  \ts{FeynRules}.%
\fn{\ts{FeynMaster} will \textit{not} run if a \t{\ts{Python}} version prior to 3 is used, and it is only guaranteed to properly work if a version of \t{\ts{Mathematica}} not older than 10.3 is used.}
As for \LaTeX, the user is required to update the package database; note also that, in the first run of \ts{FeynMaster}, some packages (like \t{feynmp-auto} and \t{breqn}) may require authorization to be installed.
Finally, for the numerical calculations, the needed files are written in the \t{\ts{Fortran}} 77 format. A \t{\ts{Fortran}} compiler is needed, as well as an instalation of \ts{LoopTools}. The programs were shown to work both with the gfortran and ifort compilers and with LoopTools-2.14 and above.

\section{Creating a new model}
\label{sec:Create}

\n \ts{FeynMaster} is a model dependent program: it cannot run without the specification of a model.
As already mentioned, one of the major simplifications of \ts{FeynMaster} 2 lies in model building: while in the previous version the specification of each model required two files, in \ts{FeynMaster} 2 only one file is required to fully define a model. 
This one file --- which we now identify as \textit{the} \ts{FeynMaster} model file --- is essentially a regular model file for \ts{FeynRules};
as such, it has the termination \t{.fr}.%
\fn{It is assumed here that the user is familiar with \ts{FeynRules} and knows how to create a \ts{FeynRules} model. If this is not the case, we refer to the \ts{FeynRules} website \url{https://feynrules.irmp.ucl.ac.be/}. Finally, it should be clear that the \ts{FeynMaster} model file can always be used as a \ts{FeynRules} model file.}
Its name is identified in the following as the \textit{internal name} of the model. This model file must be inside a folder with the same name --- that is, with the internal name of the model ---, which in turn must be inside the folder with the models for \ts{FeynMaster}.

\n \ts{FeynMaster} already comes with three models : QED, Scalar QED (SQED) and the Standard Model (SM).\fn{The SM model file is written in an arbitrary $R_{\xi}$ gauge and with the $\eta$ parameters of ref.~\cite{Romao:2012pq}, and it closely follows ref.~\cite{Denner:1991kt} for renormalization.}
These serve as prototypes, and we highly recommend them as guiding tools in the creation of a new model.
Since there are already models available, this section can be skipped in a first utilization of \ts{FeynMaster}.

\n We now describe in detail the creation of a \ts{FeynMaster} model file. As we just mentioned, this file is essentially a model for \ts{FeynRules}. Yet, it is characterized by special features: while some of the definitions and attributes are common to \FRR, some were specifically designed for \FMM, as we now show.

\vs{3mm}
\n \underline{\textbf{Lagrangian}}

\n The \ts{FeynMaster} model file must include the Lagrangian of the theory. The Lagrangian must be separated in different parts --- each of them corresponding to a different type of interaction --- and each of those parts should have a specific name: see Table \ref{tab:FRnames}.\fn{It is not mandatory to define all the 6 Lagrangian parts present in Table \ref{tab:FRnames} (for example, if the model does not have a ghost sector, there is no need to define LGhost); only, no other name besides the 6 ones specified in the right column of Table \ref{tab:FRnames} will be recognized by \ts{FeynMaster}.}
\begin{table}[!h]%
\begin{normalsize}
\normalsize
\begin{center}
\begin{tabular}
{@{\hspace{3mm}}>{\raggedright\arraybackslash}p{5cm}>{\raggedright\arraybackslash}p{2.9cm}@{\hspace{3mm}}}
\hlinewd{1.1pt}
Type of interactions & \ts{FeynRules} name \\
\hline
pure gauge & \t{LGauge} \\
fermion-gauge & \t{LFermions}\\
Yukawa & \t{LYukawa} \\
scalar-scalar and gauge-scalar & \t{LHiggs} \\
ghosts & \t{LGhost} \\
gauge fixing & \t{LGF} \\
\hlinewd{1.1pt}
\end{tabular}
\end{center}
\vspace{-5mm}
\end{normalsize}
\caption{Name of the different Lagrangian parts according to the type of interaction.}
\label{tab:FRnames}
\end{table}
\normalsize
The Lagrangian parts involving fermions (\t{LFermions} and \t{LYukawa}) can be written either in terms of Dirac fermions or Weyl fermions; in both cases, the Feynman rules will be presented for Dirac fermions.

\vs{3mm}
\n \underline{\textbf{Parameters}}

\n Parameters are defined via the usual \FR variable \t{M\$Parameters}.
Different attributes can be associated to a certain parameter; their order is irrelevant.
As in \FRR, the parameters are by default real; otherwise, one must include the attribute \t{ComplexParameter} and set it to \t{True} (i.e., \t{ComplexParameter -> True}).
The \LaTeX \, name of a parameter should be set with the attribute \t{TeXName}\fn{Which does not exist in \FR and should not be confused with the \FR attribute \t{TeX}.} and written between double commas in \LaTeX \, style (ex: \verb|TeXName -> "\\lambda_2"|).\fn{The backslash should always be doubled in \t{TeXName}, as well as in \t{TeXAntiName} (see below).}
Parameters can have indices, which are defined as in \FR (ex: \t{Indices -> \{Index[Gen], Index[Gen]\}}).%
\fn{In this example, the range of \t{Gen} must be defined in the model; for example,
\t{IndexRange[Index[Gen]]} \t{= Range[3]}.}
A numerical value can be given to a parameter through the attribute \t{NumValue} (ex: \verb|NumValue -> 0.0047|); in the case of parameters with indices, the set of values should be written as a \t{\ts{Mathematica}} list (ex: for a parameter with dimensions $2\times2$, one can set \t{NumValue -> \{0.5, 1, 1.5, 2\}}, for the matrix entries 11, 12, 21, 22, respectively).%
\fn{The attribution of numerical values is useful only in case the user wants to exploit the numerical interface of \ts{FeynMaster} --- to be explained in section \ref{sec:FC}. The SM model file includes numerical values according to \cite{Tanabashi:2018oca}.}
The renormalization rule for a parameter can be defined through the attribute \t{Renormalization}, in such a way that the rule must be written inside braces (ex: \verb|Renormalization -> {mf -> mf + dmf}|).\fn{It should be clear that it is not mandatory to define renormalization rules. That is, \FM can be used even if no renormalization is performed.
Finally, note that, if a parameter is complex, the renormalization of its complex conjugate is automatically considered.}
Finally, parameters corresponding to counterterms should be flagged by setting the attribute \t{Counterterm} to \t{True}.\fn{There can be counterterms for both parameters and fields; since counterterms themselves are parameters, they are defined in the parameters sections \t{M\$Parameters}.}

\vs{3mm}
\n \underline{\textbf{Particle classes}}

\n Particles are defined via the usual \FR variable \t{M\$ClassesDescription};
again, the order of attributes associated to a certain parameter is irrelevant.
The \t{ClassName} cannot correspond to \FM internal indices, such as \t{J1}, \t{J2}, \t{J3}, \t{J4}, \t{p1}, \t{p2}, \t{q1}, \t{q2}.
The attribute \t{SelfConjugate} should always be defined, either to \t{True} or to \t{False}.
Unphysical particles (except Weyl fermions) should be flagged with the attribute \t{Unphysical} set to \t{True}.
Every propagating particle must have a mass assignment, either to a variable or to 0 (ex: \t{Mass -> MH} or \t{Mass -> 0}).
Propagating particles should also have a \LaTeX \, name (ex: \verb|TeXName -> "c_{W^-}"|); when a particle is not its own complex conjugate, a \LaTeX \, name for the antiparticle should also be specified, through the attribute \t{TeXAntiName} (ex: \verb|TeXAntiName -> "\\bar{c_{W^-}}"|).
The decay width can be defined through \t{DecayWidth} (ex: \verb|DecayWidth -> DWZ|).%
\fn{The decay width thus defined will show up in the denominator of the propagator of the particle at stake whenever such propagator mediates the $s$-channel of a 2 $\to$ 2 scattering process. This will also automatically happen for the corresponding Goldstone boson of the particle at stake (in the case it corresponds to a gauge boson) if, in the definition of the particle corresponding to the Goldstone, the attribute \t{Goldstone} is associated to the gauge boson at stake (ex: \t{Goldstone -> Z}).}
Weyl components must be specified for Dirac fermions whenever the latter are composed of Weyl particles defined in the model (ex: \verb|WeylComponents -> {uqL,uqR}|).
Neutrinos should be flagged with the attribute \t{Neutrino} as \t{True}.%
\fn{This is relevant only for the definition of number of polarizations in decay or scattering processes. In particular, the functions \t{DecayWidth} and \t{DiffXS} (cf. Table \ref{tab:FCfunc} below) use this information.}
Finally, renormalization is defined in a way similar to that of the parameters (ex: \t{Renormalization -> \{H -> H + 1/2 dZH H\}}).%
\fn{In the case of gauge bosons, generic Lorentz indices should be included; ex: 
\t{A[mu_] -> A[mu] + 1/2 dZ3 A[mu]}. Renormalization of antiparticles is automatically included and needs not be specified by the user.}

\vs{3mm}
\n \underline{\textbf{Extra (optional) information}}

\n Besides \t{M\$Parameters} and \t{M\$ClassesDescription}, \FM allows extra optional quantities to be defined.
These are presented in Table \ref{tab:FMmodel} and for each one we present a brief description, the default value and an example. A few of comments are in order.
%
%
\begin{table}[!h]%
\begin{center}
\begin{tabular}
{@{\hspace{3mm}}
>{\raggedright\arraybackslash}p{2.7cm}
>{}p{5.4cm}
>{}p{2.5cm}
>{}p{3.2cm}
@{\hspace{3mm}}}
\hlinewd{1.1pt}
Quantity & Description & Default value & Example\\
\hline\\[-2mm]
\t{M\$ModelExtName} &  model name to be printed in the \ts{FeynMaster} final documents & \textit{(internal name)} & \t{"Standard Model"}\\[7mm]
\t{M\$FCeqs} & identities for \ts{FeynCalc} & \textit{(empty)} & \t{\{xiA->1\}}\\[7mm]
\t{M\$FCsimp} & simplifications for \ts{FeynCalc} & \textit{(empty)} & \t{\{cw^2+sw^2->1\}}\\[7mm]
\t{M\$PrMassFL} & when true, masses of propagators are extracted from the Lagrangian & \t{True} & \t{False}\\[7mm]
\t{M\$GFreno} & when true, renormalization rules are applied to the Gauge Fixing Lagrangian & \t{False} & \t{True}\\[10mm]
\t{M\$RestFile}  & restrictions file for the \ts{FeynRules} model & \textit{(empty)} & \t{"MyRest.rst"}\\[10mm]
\t{RenoPreRep} & function to rewrite the Lagrangian immediatly before renormalization & \textit{(empty)} & \textit{(cf. SM model file)}\\
\hlinewd{1.1pt}
\end{tabular}
\end{center}
\vspace{-5mm}
\caption{Optional variables for the \FM model file. See text for details.}
\label{tab:FMmodel}
\end{table}
\normalsize

\n First, \t{M\$FCeqs} should be distinguished \t{M\$FCsimp}: while both are defined in the \FM model file as lists of replacement rules, the former ends up being converted into a list of equalities for the \FCC, whereas the latter is only used as a set of replacement rules. So, considering the examples presented in Table \ref{tab:FMmodel}, the \FC notebook will have \t{xiA} defined as 1 (i.e., \t{xiA:=1}), while it will use the rule \t{cw^2+sw^2->1} to simplify the calculations (more details on section \ref{sec:FC}).

\n Second, if \t{PrMassFL} is set to \t{True}, the poles of the propagators are extracted from the Lagrangian --- i.e., they are defined as the bilinear terms of the field at stake in the Lagrangian --- and the propagator is written in the most general form. If \t{PrMassFL} is set to \t{False}, the poles will match the variable corresponding to the mass of the propagator and the propagator will be written in the Feynman gauge. While setting \t{PrMassFL} to \t{True} is certainly the most faithful way to describe the propagator, this option may bring certain difficulties: on the one hand, it requires the definition of a Gauge Fixing term for gauge boson propagators; on the other hand, the bilinear terms can be very complicated expressions.

\n Finally, \t{RenoPreRep}, if used, must be defined as a \t{\ts{Mathematica}} function with a single argument. It is applied to the Lagrangian immediatly before the renormalization process, thus allowing to rewrite it (i.e., the Lagrangian) in a more convenient way. This may be useful to obtain simpler expressions for the Feynman rules of the counterterms.

\vs{-1mm}
\section{Usage}
\label{sec:Usage}

\n Once the initial specifications are concluded (i.e., once the directories and the model are defined), \ts{FeynMaster} is ready to be used. In this section, we explain in detail how to use it. We start by showing how to edit the file that controls the \ts{FeynMaster} run. Then, after describing how to run \ts{FeynMaster}, we comment its outputs and explain how to use the notebooks we alluded to in the Introduction.

\subsection{\t{Control.m}}
\label{sec:Control}

\n As mentioned before, the \ts{FeynMaster} run is uniquely controlled from the \t{Control.m} file (which lies inside the \t{FeynMaster} folder). In this section, we explain the different components of this file.

\vs{3mm}
\n \underline{\textbf{Model selection}}

\n First, the model must be chosen; this is done by editing the name after the variable \t{model} in the beginning of \t{Control.m}. This name must correspond to the internal name of the model (the same name given to the \FM model file).

\vs{3mm}
\n \underline{\textbf{Process specification}}

\n Then, the user should start by specifying the desired process (or processes).%
\fn{
It is possible to define a sequence of processes --- that is, a series of processes to be run in a single \ts{FeynMaster} run.
To define a sequence of processes, the user must copy the lines of the first process, paste them after it, and edit them to define the second process, and repeat the same procedure for more processes (an example of a sequence of processes will be given in section \ref{sec:FullReno}).
}
One process is specified through the definition of the set of variables shown in Table \ref{tab:process}.%
\fn{As in Table \ref{tab:FMmodel}, all variables in Table \ref{tab:process} can be completely omitted from the \t{Control.m} file, in which case the default values are applied.}
\begin{table}[!h]%
\begin{center}
\begin{tabular}
{@{\hspace{3mm}}>{\raggedright\arraybackslash}p{2.5cm}>{}p{3.9cm}>{}p{2.3cm}>{}p{4.6cm}@{\hspace{3mm}}}
\hlinewd{1.1pt}
Variable & Description & Default value & Example\\
\hline\\[-2mm]
\t{inparticles} & incoming particles & \textit{(empty)} & \t{e,ebar} \\[2.5mm]
\t{outparticles} & outgoing particles & \textit{(empty)} & \t{H,Z} \\[2.5mm]
\t{loops} & number of loops & \t{0} & \t{1}\\[2.5mm]
\t{parsel} & intermediate particles & \textit{(empty)} &
\multirow{3}{2.5cm}{\\[-11mm]\t{\{avoid,Z,1,3\},\{keep,e,1,1\}}}
\\[2.5mm]
\t{factor} & quantity to factor out & 1 & \t{16 Pi^2/I} \\[2.5mm] 
\t{options} & options & \textit{(empty)} & \t{onepi}\\
\hlinewd{1.1pt}
\end{tabular}
\end{center}
\vspace{-5mm}
\caption{Variables that specify one process. See text for details.}
\label{tab:process}
\end{table}
\normalsize
As before, for each variable, a brief description, the default value and an example are presented. We now describe them in more detail.

\n \t{inparticles} and \t{outparticles}, corresponding to the incoming and outgoing particles of the process, should contain only particles defined in the \ts{FeynMaster} model. Antiparticles are defined according to the \FR convention (with the suffix \t{bar}).
Different particles should be separated by commas. Whenever both a particle and an antiparticle are considered, the particle should always be written first, and the antiparticle after it. Tadpoles are obtained by selecting a single incoming particle and no outgoing particles.

\n Concerning the \t{loops} variable, it should be clear that, whatever the number of loops, \ts{FeynMaster} will always correctly generate the amplitudes for every diagram involved, although it is only prepared to properly draw and compute diagrams with number of loops inferior to 2. The one-loop calculations are performed with the \FM function \t{OneLoopTID} (cf. Table \ref{tab:FCfunc} below).

\n \t{parsel} allows the specification of intermediate particles contributing to the process.\fn{It is similar to (and actually based on) the \t{iprop} option in \ts{QGRAF}.}
It applies not only to particles in loops, but to all intermediate particles. The selections must be written between braces and different selections should be separated by commas (see example in Table \ref{tab:process}, where two specific selections are given).
Each specific selection contains four arguments: the first should either be \t{avoid} or \t{keep}, the second should correspond to a particle of the model,\fn{Care should be taken not to select antiparticles, but only particles. This is because the propagator in \ts{FeynMaster} is defined through the particle, and not the antiparticle.} and the last two should be non-negative integer numbers such that the second is not smaller than the first.
We illustrate how it works by considering the example in Table \ref{tab:process}: \t{\{avoid,Z,1,3\}} discards all the diagrams with number of \t{Z} propagators between 1 and 3, while \t{\{keep,e,1,1\}} keeps only diagrams with number of \t{e} propagators between 1 and 1 (i.e., exactly equal to 1).

\n \t{factor} is a number, written in \t{\ts{Mathematica}} style that is to be factored out in the final calculations (more details on section \ref{sec:FC}).

\n \t{options} allows all the \ts{QGRAF} options, as well as a couple of specific option for \FMM --- \t{ugauge}, which performs the calculation in the unitary gauge, and \t{noPVauto}, which employs the option \t{PaVeAutoReduce -> False} in \t{TID}.%
\fn{\label{note:TID}
As shall be seen, loop integrals in \FM are performed with the function \t{OneLoopTID}, which uses the \FC function \t{TID}. By default, \t{OneLoopTID} uses \t{TID} with the option \t{PaVeAutoReduce -> True}. This simplifies some special cases of Passarino-Veltman functions. In some cases, however, it may be relevant to perform loop integrals in a more expanded way, in which case the option \t{PaVeAutoReduce -> False} must be used by selecting the option \t{noPVauto} in \t{Control.m}.}
. In Table \ref{tab:Qoptions}, besides these, we present some of the \ts{QGRAF} options; for each one, a brief description and the converse option are presented.

\begin{table}[!h]%
\begin{normalsize}
\normalsize
\begin{center}
\begin{threeparttable}
\begin{tabular}
{@{\hspace{3mm}}
>{\raggedright\arraybackslash}p{2.5cm}
>{\raggedright\arraybackslash}p{7.2cm}
>{\raggedright\arraybackslash}p{2.8cm}
@{\hspace{3mm}}}
\hlinewd{1.1pt}
Option & Description & Converse option \\
\hline\\[-1.5mm]
\t{ugauge} & unitary gauge & \textit{(none)}\\[3mm]
\t{noPVauto} & employs \t{PaVeAutoReduce -> False} in \t{TID} & \textit{(none)}\\[3mm]
\t{onepi} & 1-particle irreducible diagrams only\tnote{$\star$} & \t{onepr}\\[3mm]
\t{onshell} & no self-energy insertions on the external lines\tnote{$\star$} & \t{offshell}\\[3mm]
\t{nosigma} & no self-energy insertions (nowhere)\tnote{$\star$} & \t{sigma}\\[3mm]
\t{nosnail} & no snails (i.e., tadpoles or a collapsed tadpoles)\tnote{$\star$} & \t{snail}\\[3mm]
\t{notadpole} & no tadpole insertions, i.e. no 1-point insertions\tnote{$\star$} & \t{tadpole}\\[3mm]
\t{simple} & at most one propagator connecting any two different vertices, and no propagator starting and ending at the same vertex\tnote{$\star$} & \t{notsimple}\\
\hlinewd{1.1pt}
\end{tabular}
\begin{tablenotes}
{\small
\item[$\star$] Taken from the \ts{QGRAF} manual; please consult it for more information.
}
\end{tablenotes}
\end{threeparttable}
\end{center}
\vspace{-5mm}
\end{normalsize}
\vs{-1mm}
\caption{Options for \t{Control.m}.}
\label{tab:Qoptions}
\end{table}

\vs{3mm}
\n \underline{\textbf{\FM logical variables}}

\n Once the model and the process (or sequence of processes) are specified, the user must select the logical value of the 9 variables present in the end of \t{Control.m}. As logical variables, they admit only the values \t{True} (or, alternatively, \t{T} or \t{t}) and \t{False} (or, alternatively, \t{F} or \t{f}).
We describe them in Table \ref{tab:sel}. 
\begin{table}[!h]%
\begin{normalsize}
\normalsize
\begin{center}
\begin{tabular}
{@{\hspace{3mm}}>{\raggedright\arraybackslash}p{2.8cm}>{\raggedright\arraybackslash}p{7.2cm}@{\hspace{3mm}}}
\hlinewd{1.1pt}
Variable & Effect (when chosen as \t{True}) \\
\hline\\[-1.5mm]
\t{FRinterLogic} & establish an interface with \ts{FeynRules} \\[2.5mm] 
\t{RenoLogic} & perform renormalization \\[2.5mm]
\t{Draw} & draw and print the Feynman diagrams \\[2.5mm]
\t{Comp} & compute the final expressions \\[2.5mm]
\ \ \t{FinLogic} & print the final result of each diagram \\[2.5mm]
\ \ \t{DivLogic} & print the UV divergent part of each diagram \\[2.5mm]
\t{SumLogic} & compute the sum of the expressions\\[2.5mm]
\t{MoCoLogic} & apply momentum conservation \\[2.5mm]
\t{LoSpinors} & include spinors \\[2.5mm]
\hlinewd{1.1pt}
\end{tabular}
\end{center}
\vspace{-5mm}
\end{normalsize}
\caption{Logical variables of \t{Control.m}. See text for details.}
\label{tab:sel}
\end{table}
\normalsize
Some remarks are in order, concerning the effect of these variables when set to \t{True}.

\n \t{FRinterLogic}, by establishing an interface with \ts{FeynRules}, performs several tasks. First, it runs \ts{FeynRules} (for the model selected in the initial variable \t{model}), prints the complete tree-level Feynman rules of the model in a PDF file and opens this file.\fn{\label{nt:1st}The results are automatically written; this is especially challenging when it comes to (automatically) breaking the lines in a long equation. This challenge is in general surpassed with the \LaTeX\, \t{breqn} package, which \ts{FeynMaster} employs. However, \t{breqn} is not able to break a line whenever the point where the line is to be broken is surrounded by three or more parentheses; in those cases, unfortunately, the lines in the \ts{FeynMaster} PDF outputs simply go out of the screen. For documentation on the \t{breqn} package, cf.\url{https://www.ctan.org/pkg/breqn}.}
Second, it generates the \ts{QGRAF} model file, a crucial element in the generation of Feynman diagrams. Third, it generates the complete tree-level Feynman rules in \ts{FeynCalc} style, which will play a decisive role in all the calculations. Finally, it generates a \t{\ts{Mathematica}} notebook specifically designed to run \ts{FeynRules} --- hereafter dubbed the \ts{FeynRules} notebook. This notebook is useful in case the user wants to have control over the generation of Feynman rules, and is the subject of section \ref{sec:FR}.

\n Note that, even if all logical variables are set to \t{False}, \ts{FeynMaster} always performs some actions. First, it
generates a \t{\ts{Mathematica}} notebook specifically designed to run \ts{FeynCalc} --- hereafter dubbed the \ts{FeynCalc} notebook. This notebook is very useful should the user want to have control over calculations, and is the subject of section \ref{sec:FC}. Second, once there is a \ts{QGRAF} model, \ts{FeynMaster} always runs \ts{QGRAF}, which writes in a symbolic form the total diagrams that contribute to the process at stake --- the same process which was specified through the variables in Table \ref{tab:process}. Finally, \ts{FeynMaster} takes the \ts{QGRAF} output and writes the amplitude for each diagram in a file that the \ts{FeynCalc} notebook shall have access to.

\n \t{RenoLogic} concerns the renormalization of the model. If \t{FRinterLogic} is set to \t{True}, \t{RenoLogic} prints the complete set of Feynman rules for the counterterms interactions in a PDF file and opens this file; moreover, it stores those interactions in a file which the \FC notebook shall have access to. A second important feature of \t{RenoLogic} is described below, in the context of the \t{Comp} variable.

\n \t{Draw} takes the \ts{QGRAF} output, draws the Feynman diagrams in a \LaTeX \,file, prints them in a PDF file and opens this file. This operation is achieved with the help of \t{feynmf} \cite{feynmf}, a \LaTeX \,package to draw Feynman diagrams. Since the diagrams are written in a \LaTeX \,file, they can not only be edited by the user, but also directly copied to the \LaTeX \,file of the user's paper.\fn{As already suggested, \t{Draw} is at present only guaranteed to properly draw the diagrams up to one-loop. Moreover, diagrams with more than two particles in the initial or final states, as well as some reducible diagrams, are also not warranted.}

\n \t{Comp} computes the final expressions using \ts{FeynCalc} and stores them in a file.%
\fn{By `final expressions' we mean the analytical expressions for the diagrams; these are written in terms of Passarino--Veltman integrals in case the number of loops equals 1; more details on section \ref{sec:FC}. It is normal that the warning `\textit{front end is not available}' shows up when \t{Comp} is set to \t{True}.}

\n \t{FinLogic} and \t{DivLogic} are nothing but options for \t{Comp}, so that their value is only relevant if \t{Comp} is set to \t{True}.
If at least one of them is set to \t{True}, a PDF file is generated and opened: when \t{FinLogic} is selected, the file includes the (total) final analytical expression for each diagram; when \t{DivLogic} is selected, it includes the analytical expression for the UV divergent part of each diagram.\fn{The limitation we alluded to in note \ref{nt:1st} applies here too.}

\n At this point, we should clarify the difference between UV divergences and infrared (IR) divergences. It is well known that, while the former are in general present in loop integrals, the latter can only show up when there is a massless particle running inside the loop (in which case the IR divergence comes from the integration region near $k^2=0$, with $k$ the loop momentum). In the present version of \ts{FeynMaster}, we restrict the treatment of divergences to the UV ones. Indeed, we assume that the IR divergences can be regulated by giving the massless particle a fake mass --- which one shall eventually be able to set to zero in physical processes, after considering real emission graphs. With that assumption, IR divergences will never show up explicitly (only implicitly through the fake mass). In the following, unless in potentially dubious statements, we will stop writing UV explicitly: it is assumed that, whenever we mention divergences, we shall be referring to UV divergences.

\n \t{SumLogic} is relevant both when \t{Comp} is \t{True} and when it is \t{False}. In the first case, it calculates the sum of the analytical expressions (when the PDF file with the analytical expressions is generated, \t{SumLogic} prints their sum).%
\fn{More specifically: if \t{FinLogic} is \t{True}, \t{SumLogic} includes in the PDF file the sum of the total final expressions; if \t{DivLogic} is \t{True}, it includes the sum of the expressions for the divergent parts; if both are \t{True}, it includes both the sum of the total expressions and the sum of the expressions for the divergent parts.}
The situation where \t{SumLogic} is \t{True} and \t{Comp} is \t{False} shall be described in section \ref{sec:FC}.

\n We now explain the effect of \t{RenoLogic} when \t{Comp} is set to \t{True}.
In case the user defined a single process in \t{Control.m}, \t{RenoLogic} causes \ts{FeynMaster} to look for counterterms that might absorb the divergences of the process at stake, and to calculate those counterterms in $\overline{\text{MS}}$.%
\fn{That is, calculates them in such a way that the counterterms are precisely equal to the divergent part they absorb (except for the $\ln(4\pi)$ and the Euler-Mascheroni  constant $\gamma$, which are also absorbed in the $\overline{\text{MS}}$ scheme). By `calculating' we mean here writing the analytical expression.}
Such counterterms are then stored in a file (\t{CTfin.m}, described below) and,
should the PDF file with the analytical expressions be generated, printed in such file.
In case the user defined a sequence of processes, the subsequently computed counterterms are added to \t{CTfin.m}; however, what is particularly special about the sequence is that, for a certain process of the sequence, \ts{FeynMaster} will compute the counterterms by making use of the counterterms already computed in the previous processes.%
\fn{This is, in fact, the major advantage of writing a series of processes in a single \ts{FeynMaster} run (as opposed to one process per run); indeed, this feature is not possible with one process per run, since \FM rewrites \t{CTfin.m} for the model at stake everytime \t{Comp} and \t{RenoLogic} are both set to \t{True}.}
In the end of the run, \t{CTfin.m} contains all the counterterms that were computed (in $\overline{\text{MS}}$) to absorb the divergences of the processes of the sequence. In this way, and by choosing an appropriate sequence of processes, it is possible to automatically renormalize the whole model in $\overline{\text{MS}}$ with a single \ts{FeynMaster} run.

\n As \t{SumLogic} described above, the last two logical variables of Table \ref{tab:sel} --- \t{MoCoLogic} and \t{LoSpinors} --- are relevant both when \t{Comp} is \t{True} and when it is \t{False}. \t{LoSpinors} --- only relevant with external fermions --- is actually prior to any calculation, since it modifies the amplitudes themselves, including spinors in them. When \t{Comp} is \t{True}, \t{MoCoLogic} is such that the calculations use momentum conservation. The situation where \t{MoCoLogic} is \t{True} and \t{Comp} is \t{False} is also postponed to section \ref{sec:FC}.

\vs{3mm}
\n \underline{\textbf{Tips}}

\n When using \ts{FeynMaster}, it can be convenient to have access to some tips in order to properly fill in \t{Control.m}. Several such tips are available: in the command line, going to the directory corresponding to the \t{FeynMaster} folder, one can run the \t{RUN-\ts{FeynMaster}} batch file with an argument after it. If this argument is simply \textit{help}, then several instructions about the use of \t{Control.m} are printed in the command line.
If, instead, the argument corresponds to internal name of a \t{FeynMaster} model, all the particles of the model are printed.%
\fn{Antiparticles are not shown, and are obtained simply by adding \t{bar} after the name of the corresponding particle.}
If the argument is neither \textit{help} nor the name of a \t{FeynMaster} model, all the particles of the model currently selected in \t{Control.m} are printed.

\subsection{Run}
\label{sec:Run}

\n After editing the \t{Control.m} file, everything is set. To run \ts{FeynMaster}, just run the \t{RUN-\ts{FeynMaster}} batch file inside the \t{FeynMaster} folder (with no arguments after it).\fn{Care should be taken not to run \ts{FeynMaster} when the relevant notebooks are open. More specifically, if \t{FRinterLogic} is set to \t{True}, and if the \ts{FeynRules} notebook created for the process at stake already exists, this notebook cannot be open during the run; in the same way, if \t{Comp} is set to \t{True}, and if the \ts{FeynCalc} notebook designed for the process at stake already exists, such notebook cannot be open during the run.}

\subsection{Outputs}
\label{sec:Output}

\n Depending on the logical value of the variables of Table \ref{tab:sel}, \ts{FeynMaster} can have different outputs. We now list the total set of outputs, assuming that all those variables are set to \t{True}.%
\fn{Actually, when \t{Comp} and \t{RenoLogic} are both \t{True} and there are external fermions, \t{LoSpinors} should be \t{False}. This is irrelevant for what follows, since \t{LoSpinors} has no influence on the outputs as a whole.}
First, in the directory where the \FM model is, two files are generated: the \ts{FeynRules} notebook (\t{Notebook.nb}) and an auxiliary file for it (\t{PreControl.m}). Second, the \ts{QGRAF} model with the internal name of the model is created, and placed inside the folder with the models for \ts{QGRAF} (cf. note \ref{note:instal}); besides, a file named \t{last-output} (with the last output from QGRAF) is created inside the directory corresponding to \ts{QGRAF}. Then, if it does not exist yet, a directory with the internal name of the model is created inside the directory for the \FM output. Inside it, and if they do not exist yet, three directories are created, \t{Counterterms}, \t{FeynmanRules} and \t{Processes}, which we now describe.

\n \t{Counterterms} contains one folder, \t{TeXs-drawing}, and two files, \t{CTini.m} and \t{CTfin.m}. \t{TeXs-drawing} is where the PDF file with the complete set of Feynman rules for the counterterms interactions is stored, as well as the \LaTeX \, file that creates it. \t{CTini.m} is the file which the \ts{FeynCalc} notebook has access to and where the Feynman rules for the counterterms interactions are stored. \t{CTfin.m}, in turn, is the aforementioned file containing the counterterms that were computed (in $\overline{\text{MS}}$) to absorb the divergences of the processes of the sequence at stake.

\n \t{FeynmanRules}, besides several auxiliary files to be used in the \ts{FeynCalc} notebook, contains yet another \t{TeXs-drawing} folder, where the PDF file with the complete  set of Feynman rules for the
tree-level interactions is stored, as well as the \LaTeX \, file that creates it.

\n \t{Processes} contains a folder for each of the different processes studied. These folders are named with the index (in the sequence of processes) corresponding to the process at stake, as well as with a string containing the names of the incoming and the outgoing particles joined together.
Inside each folder, there are three other folders, \t{Lists}, \t{TeXs-drawing} and \t{TeXs-expressions}, as well as three files, \t{Amplitudes.m}, \t{Helper.m} and the \ts{FeynCalc} notebook, \t{Notebook.nb}.
In order:
\t{Lists} contains files where the analytical expressions for the process at staked are stored (more details below);
\t{TeXs-drawing} contains the PDF file with the printed Feynman diagrams, as well as the \LaTeX \,file that creates it;
\t{TeXs-expressions} contains the PDF file with the printed expressions, as well as the \LaTeX \,file that creates it;
\t{Amplitudes.m} contains the amplitudes for the diagrams (written in \ts{FeynCalc} style);
\t{Helper.m} is an auxiliary file for the \ts{FeynCalc} notebook.

\n Finally, recall that, in case there is already a QGRAF model, \ts{FeynMaster} will run even if all variables of Table \ref{tab:sel} are set to \t{False}. This is relevant since it generates not only the QGRAF output (\t{last-output}), but also the folder (or folders) for the specific process (or processes) selected, containing the files described above.\fn{While the QGRAF output is not overwritten when QGRAF is run on its own, it is overwritten when QGRAF is run inside \ts{FeynMaster}.}

\subsection{The notebooks}
\label{sec:NB}

\n As previously mentioned, a major advantage of \ts{FeynMaster} is its hybrid character concerning automatization. Indeed, not only does it automatically generate the results, but it also allows the user to handle them. This is realized due to the automatic creation of the \ts{FeynRules} notebook and the \ts{FeynCalc} notebook. We now describe them in detail.

\subsubsection{The \ts{FeynRules} notebook}
\label{sec:FR}

\n We mentioned in section \ref{sec:Control} that, when \ts{FeynMaster} is run with the logical variable \t{FRinterLogic} set to \t{True}, the \ts{FeynRules} notebook \t{Notebook.nb} is automatically created in the directory where the \FM model is. By running it, the user can access the vertices for the different Lagrangian parts, according to Table \ref{tab:FRverts}.\fn{The run will generate several \ts{FeynMaster} internal files, among which is \t{built-model}, the \ts{QGRAF} model file.}
\begin{table}[!h]%
\begin{normalsize}
\normalsize
\begin{center}
\begin{tabular}
{@{\hspace{3mm}}>{\raggedright\arraybackslash}p{3.2cm}>{\raggedright\arraybackslash}p{3.8cm}@{\hspace{3mm}}}
\hlinewd{1.1pt}
Lagrangian part & vertices \\
\hline
\t{LGauge} & \t{vertsGauge} \\
\t{LFermions} & \t{vertsFermionsFlavor} \\
\t{LYukawa} & \t{vertsYukawa} \\
\t{LHiggs} & \t{vertsHiggs} \\
\t{LGhost} & \t{vertsGhosts} \\
\hlinewd{1.1pt}
\end{tabular}
\end{center}
\vspace{-5mm}
\end{normalsize}
\caption{Names of the different vertices according to the Lagrangian part (compare with Table \ref{tab:FRnames}).}
\label{tab:FRverts}
\end{table}
\normalsize
Besides the usual \ts{FeynRules} instructions, two useful functions --- \t{GetCT} and \t{MyTeXForm} --- are available.
\t{GetCT} is a function that, for a certain Lagrangian piece given as argument, yields the Feynman rules for the respective counterterms.%
\fn{In order to clarify the particles involved in the Feynman rule at stake, each term includes a factor $\eta_i$ for each particle $i$; for example, a term in the output of \t{GetCT} including \t{{$\eta$A}^2 $\eta$WP $\eta$WPbar} contributes to the Feynman rule for the counterterm of \t{A}, \t{A}, \t{WP}, \t{WPbar}.}
\t{MyTeXForm} is \ts{FeynMaster}'s version of \t{\ts{Mathematica}}'s \t{TeXForm}; it is a function that uses \t{\ts{Python}} (as well as inner \ts{FeynMaster} information concerning the \LaTeX \,form of the parameters of the model) to write expressions in a proper \LaTeX \,form.%
\fn{\t{MyTeXForm} prints the \LaTeX \,form of the expression at stake not only on the screen, but also in an external file named \t{MyTeXForm-last-output.tex} in the directory where the notebook lies.}

\subsubsection{The \ts{FeynCalc} notebook}
\label{sec:FC}

\n Whenever \ts{FeynMaster} is run, and independently of the logical values of the variables of Table \ref{tab:sel}, the \ts{FeynCalc} notebook is automatically created. This notebook, as already the \ts{FeynRules} one just described, is totally ready-to-use: the user does not have to define directories, nor import files, nor change conventions.
Just by running the notebook, there is immediate access to a whole set of results: not only to some basic elements --- such as the total Feynman rules for the model and amplitudes for the diagrams ---, but also to 
the totality of the results obtained should the \t{Comp} logical variable had been turned on. This last feature is made possible in \FM 2 due to the creation of lists with analytical expressions. Indeed, when \t{Comp} is set to true, \FM now stores the analytical expressions inside the aforementioned folder \t{Lists}, in such a way that, after the run, the \ts{FeynCalc} notebook has immediate access to those expressions (more details below). 

\n But this is just part of the flexibility involved in the \ts{FeynCalc} notebook. In fact, since all the referred results are written in a \t{\ts{Mathematica}} notebook, the user has great control over them, as he or she can operate algebraically on them, or select part of them, or print them into files, etc.
Moreover, since the \ts{FeynCalc} package is loaded, and since all the results are written in a \ts{FeynCalc}-readable style, the control at stake is even greater, for the user can apply all the useful tools of that package: operate on the Dirac algebra, perform contractions, solve loop integrals, etc.\fn{In the following, we assume the user to be familiar with \ts{FeynCalc}. For more informations, consult the \ts{FeynCalc} website: \url{https://feyncalc.github.io/}.} 

\n We now present some useful features introduced by \ts{FeynMaster} in the \ts{FeynCalc} notebook. We start with the variables related to the analytical expressions for the Feynman diagrams: see Table \ref{tab:FCvars}.
\begin{table}[!h]%
\begin{normalsize}
\normalsize
\begin{center}
\begin{tabular}
{@{\hspace{3mm}}>{\raggedright\arraybackslash}p{1.8cm}>{\raggedright\arraybackslash}p{8.0cm}@{\hspace{3mm}}}
\hlinewd{1.1pt}
Variable & Meaning \\
\hline\\[-1.5mm]
\t{amp} & list with all the amplitudes \\[2.5mm]
\t{amp}\textit{i} & amplitude for diagram \textit{i} \\[2.5mm]
\t{res} & list with all the final expressions \\[2.5mm]
\t{res}\textit{i} & final expression for diagram \textit{i} \\[2.5mm]
\t{resD} & list with all the expressions for the divergent parts \\[2.5mm]
\t{resD}\textit{i} & expression for the divergent part of diagram \textit{i} \\[2.5mm]
\t{restot} & sum of all the final expressions \\[2.5mm]
\t{resDtot} & sum of all the expressions for the divergent parts \\[2.5mm]
\hlinewd{1.1pt}
\end{tabular}
\end{center}
\vspace{-5mm}
\end{normalsize}
\caption{Useful variables concerning expressions for the diagrams. See text for details.}
\label{tab:FCvars}
\end{table}
\normalsize
We should clarify the meaning of final expression, corresponding to the \t{res} list: for a certain index $j$, \t{res}\textit{j} (or \t{res[[j]]}) takes the amplitude \t{amp}\textit{j}, divides it by $i$, rewrites the loop integral in terms of Passarino--Veltman integrals (in case it is a one-loop process), writes it in 4 dimensions --- including possible finite parts coming from this conversion%
\fn{As is well known, in the {dimensional regularization scheme}, the infinities are tamed by changing the dimensions of the integrals from 4 to $d$, in such a way that the divergences are regulated by the parameter $\epsilon = 4 - d$. When solving the integrals in terms of Passarino--Veltman integrals, the result will in general depend explicitly on the dimension $d$, as well as on the Passarino--Veltman integrals themselves --- which usually diverge, with divergence proportional to $1/\epsilon$. But since $d = 4-\epsilon$, there will in general be finite terms (order $\epsilon^0$) coming from the product between $d$ and the divergent parts in the Passarino--Veltman integrals. Hence, when converting the result back to 4 dimensions (since the final result is written in 4 dimensions), one cannot forget to include such terms. {Finally, recall that IR divergences will never show up explicitly if the potentially IR divergent integrals are tamed by giving the massless particle a fake mass.}\label{note:DimReg}}
--- and factorizes the previously selected \t{factor}.
Note that, due to the division by $i$, the final expressions --- that is, \t{res} --- correspond to $\mathcal{M}$, and not to $i \mathcal{M}$.
The divergent parts are written in terms of the variable \t{div}, defined as:%
\fn{In the PDF file with the printed expressions, we change the name \t{div} to $\omega_{\epsilon}$.}
\begin{equation*}
\t{div} = \dfrac{1}{2} \left( \dfrac{2}{\epsilon} - \gamma + \ln 4\pi \right).
\end{equation*}

\n Let us now consider the details related to the \t{Lists} folder. When \t{Comp} is set to \t{True}, the lists \t{res} and \t{resD} are calculated and stored in \t{Lists} in the files \t{res.in} and \t{resD.in}, respectively. If, in addition, \t{SumLogic} is set to \t{True}, \t{restot} and \t{resDtot} are also calculated and stored in \t{Lists} in the files \t{restot.in} and \t{resDtot.in}, respectively.
The access of the \ts{FeynCalc} notebook to these files depends on the variable \t{compNwrite}, which is visibly defined in the middle of the notebook: when it is \t{False}, the notebook will load the files stored in \t{Lists} (should they exist); when it is \t{True}, it will calculate them anew.
By default, \t{compNwrite} is \t{False}, which means that, after having been calculated with \t{Comp}, the analytical expressions are by default immediatly accessible to the \ts{FeynCalc} notebook.

\n Next, we consider useful functions to manipulate the results: see Table \ref{tab:FCfunc}.
\begin{table}[!h]%
\begin{normalsize}
\normalsize
\begin{center}
\begin{threeparttable}
\begin{tabular}
{@{\hspace{3mm}}
>{\raggedright\arraybackslash}p{2.3cm}
>{\raggedright\arraybackslash}p{11.0cm}@{\hspace{3mm}}}
\hlinewd{1.1pt}
Function & Action \\
\hline\\[-2.5mm]
\t{ChangeTo4} & change to 4 dimensions\\[2.5mm]
\t{DecayWidth} & calculate the decay width\tnote{$\star$}\\[2.5mm]
\t{DiffXS} & calculate the differential cross section\tnote{$\dagger$}\\[2.5mm]
\t{GetDirac} & yield the Dirac structures as a list\\[2.5mm]
\t{GetDiv} & get the divergent part of an expression \\[2.5mm]
\t{GetFinite} & get the finite part of an expression in dimensional regularization\tnote{$\ddagger$}\\[2.5mm]
\t{FacToDecay} & rewrite expressions (with form factors) for decays\tnote{$\star$} \tnote{$\diamond$}\\[2.5mm]
\t{FCtoFT} & convert expressions to \t{\ts{Fortran}} \\[2.5mm]
\t{MyTeXForm} & write expressions in a proper \LaTeX \,form  \\[2.5mm]
\t{MyPaVeReduce} & apply \ts{FeynCalc}'s \t{PaVeReduce} and convert to 4 dimensions\\[2.5mm]
\t{OneLoopTID} & solve one-loop integrals with \FC TID decomposition method\tnote{$\#$}\\[2.5mm]
\t{RepDirac} & replace the Dirac structures with elements \t{ME[j]}\\[2.5mm]
\t{TakeReal} & applies the operator Re to the Passarino-Veltman functions\\[2.5mm]
\t{TrG5} & calculate the trace for expressions with $\gamma_5$\\[0.5mm]
\hlinewd{1.1pt}
\end{tabular}
\begin{tablenotes}
{\small
\item[$\star$] Only applicable to processes with 1 incoming and 2 or 3 outgoing particles.
\item[$\dagger$] Only applicable to processes with 2 incoming and 2 or 3 outgoing particles.
\item[$\ddagger$] Cf. note \ref{note:DimReg}.
\item[$\diamond$] Not yet available for three external gauge bosons.
\item[$\#$] Note the arguments: \t{OneLoopTID(k,amp)}, with \t{k} the loop momentum and \t{amp} the amplitude. Cf. also note \ref{note:TID}.
}
\end{tablenotes}
\end{threeparttable}
\end{center}
\vspace{-5mm}
\end{normalsize}
\vs{-1mm}
\caption{Useful functions. See text for details.}
\label{tab:FCfunc}
\end{table}
\normalsize
We now explain some of them in more detail (note that a description of a certain function \t{Func} shows up in the notebook by writing \t{?Func}).

\n The functions \t{DecayWidth} and \t{DiffXS} have now been extended up to three outgoing particles. For the case of two body decays, the result is written solely in terms of masses; for \t{DiffXS} with two outgoing particles, the result is written also in terms of the
center of momentum energy \t{S}, as well of the scattering angle \t{Theta}. For the case of three outgoing particles the result is written in terms of the invariant scalar products,
\texttt{p0p0,p0q1,p0q2,p0q3,q1q2,q1q3,q2q3} for the decay and 
\texttt{p1p2,p1q1,p1q2,p1q3,p2q1},\texttt{p2q2},\texttt{p2q3},
\linebreak
\texttt{q1q2},\texttt{q1q3},\texttt{q2q3} for the
differential cross section. 
The argument of these functions must be a list. For the two-bodies decays, the list is either the amplitude, expr=\{res1\}, or a list with two elements, (e.g., DecayWidth[\{E0, E1\}]). In the first case, the result is proportional to |res1|^2, while in the second case the result is proportional to 2 Re [$\mathcal{E}_0$ $\mathcal{E}_1^*$]. For the case of three outgoing particles, the argument of either function must be a list of all diagrams contributing to the process. By using the function \t{FCtoFT} on the result, all necessary \t{\ts{Fortran}} files will be written in a subdirectory called \t{newfiles}, where the relevant integrations can be made.%
\fn{\label{note:complex}Both \t{DecayWidth} and \t{DiffXS} use the \ts{FeynCalc} function \t{ComplexConjugate} to perform the complex conjugate of expressions. This function assumes by default that all parameters are real; therefore, if there are complex parameters in the expression which \t{DecayWidth} or \t{DiffXS} are to be applied to, those complex parameters should be inserted in a list (say, \t{complexlist}), and one should write \t{SetOptions[ComplexConjugate, Conjugate -> complexlist]} before applying \t{DecayWidth} or \t{DiffXS}.}

\n \t{GetDirac} and \t{RepDirac} are useful functions to handle expressions with external fermions. The former yields a list containing the different Dirac structures of the expression given as argument; the latter replaces the Dirac structures of the expression given as argument with non-matricial elements \t{ME[j]} --- that is, the first Dirac structure is replaced by \t{ME[1]}, the second by \t{ME[2]}, and so on ---, thus simplying the manipulation of the expression. \t{RepDirac} admits a second (optional) argument, consisting of a list of Dirac structures (ex: \t{\{GA[p1],GS[q1]\})}; this can be relevant to establish a correspondance between \t{ME[j]} elements and Dirac structures. Indeed, such correspondance can obey one of two rules:
either only one argument is given to \t{RepDirac}, in which case the \t{ME[j]} elements follow the order of the Dirac structures yielded by \t{GetDirac}, or a list of Dirac structures is given as a second argument of \t{RepDirac}, in which case the \t{ME[j]} elements follow the order of that list.

\n \t{GetDiv} yields the UV divergent part of an expression; care should be taken if IR divergences are not regulated via a fake mass, for in that case, although they can show up as poles of the Passarino--Veltman functions, they will not be detected by \t{GetDiv}. Moreover, \t{GetDiv} only yields the UV divergent part of the Passarino--Veltman functions \cite{Denner:2005nn,tca} that \ts{FeynCalc} and \ts{FeynMaster} can handle --- namely, integrals whose power of the loop momenta in the numerator is at most 3 (except for the $D$ functions, where we extended \ts{FeynCalc} up to the fourth power of the loop momenta).

\n \t{FacToDecay} simplifies expressions to render the calculation of the decay width easier. It yields a two-element list: the first element is the expression given as argument, but rewritten in terms of form factors; the second is a list of replacements, associating to each form factor the corresponding analytical expression (written using the kinematics of the process).%
\fn{The form factors are by default defined as complex parameters in the options of the \ts{FeynCalc} function \t{ComplexConjugate}. This is relevant to calculate the DecayWidth (recall note \ref{note:complex}).}

\n \t{FCtoFT} is the function that allows the numerical interface of \ts{FeynMaster};  when applied to an expression \t{exp}, it generates a directory named \t{newfiles}, inside which there are at least five files: \t{MainFT.F}, \t{Myexp.F}, \t{MyVariables.h}, \t{MyParameters.h} and \t{Makefile-template}.%
\fn{Here, and in what follows, \t{exp} (as in \t{Myexp.F}) corresponds to the name of the expression.}
The first one, \t{MainFT.F}, is the beginning of a main \t{\ts{Fortran}} program, which must be completed according to the user's will.
\t{MainFT.F} calls the function \t{Myexp}, which is the \t{\ts{Fortran}} version of the expression \t{FCtoFT} was applied to, and which is written in the file \t{Myexp.F}.%
\fn{\label{note:extra}According to the size of the expression at stake, and in order to render the compilation faster, one or several files named \t{MyAux_exp_i.F} (with \t{i} corresponding to the index of the file) may be generated; these files are consistently and automatically called by  \t{Myexp.F}.}
In turn, \t{MyVariables.h} and \t{MyParameters.h} respectively contain the variables and the numerical values associated to the different parameters in the \ts{FeynMaster} model file. Finally, \t{Makefile-template} consists of a skeleton of a makefile.%
\fn{Whenever \t{MyAux_exp_i.F} files are generated (see note \ref{note:extra}), \t{Makefile-template} automatically includes those files, so that the compilation becomes easier.}
In the case of cross sections, the integration routine \t{IntGauss.f} is also generated and called in \t{MainFT.F}, which in this case is completed with a concrete example.%
\fn{\label{note:FCtoFT}
By default, the Passarino-Veltman functions will be converted as complex; this can be modified by defining a new variable in the \ts{FeynCalc} notebook \t{retil}, and define it as \t{True} (i.e. \t{retil=True}).
\t{FCtoFT} admits a second (optional) argument, as we now explain. Suppose that \t{exp} is a complicated expression, but such that it can be written in a simple way using three form factors, \t{F1}, \t{F2} and \t{F3}; that is, \t{exp = f(F1,F2,F3)}, where \t{f} is a simple function, while \t{F1}, \t{F2} and \t{F3}
correspond to complicated expressions. In cases like this, it is convenient to write \t{FCtoFT} with two arguments: the first one is the expression \t{exp}, but written in terms of auxiliary variables \t{F1aux}, \t{F2aux} and \t{F3aux} instead of the complicated expressions \t{F1}, \t{F2} and \t{F3} (that is,
\t{exp = f(F1aux,F2aux,F3aux)}); the second argument is a list of the replacements between the auxiliary variables and their corresponding form
factor (in this example, \t{\{F1aux -> F1, F2aux -> F2, F3aux -> F3\}}).}

\n \t{MyTeXForm} is the same function as the one described in section \ref{sec:FR}. 

\n \t{MyPaVeReduce} is \ts{FeynMaster}'s version of \ts{FeynCalc}'s \t{PaVeReduce}; it applies \t{PaVeReduce} and writes the result in 4 dimensions --- again, not without including possible finite parts coming from this conversion.

\n Finally, we consider \t{TrG5}. Since \ts{FeynMaster} is prepared to compute divergent integrals --- and, more specifically, to compute them via dimensional regularization ---, it defines Dirac and Lorentz structures (like $g^{\mu\nu}$ or $\gamma^{\mu}$) in dimension $d$, not in dimension 4. However, the definition of $\gamma_5$ in dimension $d$ is not trivial, as chiral fermions are a property of four dimensions. In fact, the treatment of $\gamma_5$ in dimensional regularization is still an open problem (see e.g. refs. \cite{Akyeampong:1973xi,Chanowitz:1979zu,Barroso:1990ti,Kreimer:1993bh,Jegerlehner:2000dz,Greiner:2002ui,Zerf:2019ynn}). By default, \ts{FeynMaster} assumes the so-called naive dimensional regularization scheme~\cite{Jegerlehner:2000dz}, which takes the relation $\{\gamma_5,\gamma^{\mu}\}=0$ to be valid in dimension $d$. This naive approach is applied both when the loop diagram has external fermions, and when it has only inner fermions (forming a closed loop). In the second case, in order to calculate the corresponding trace, \ts{FeynMaster} uses \t{TrG5}. This function starts by separating the expression it applies to in two terms: one with $\gamma_5$, another without $\gamma_5$. It then computes the trace of the former in $4$ dimensions, while keeping the dimension of the latter in its default value $d$.\fn{It is a matter of course that the calculation of the term with $\gamma_5$ in dimension 4 can only be an issue when the integral multiplying it is divergent. This is simply because one does not need to regularize an integral that is not divergent. In particular, there is no need to use dimensional regularization for a finite integral, so that all the calculations can be made in dimension 4. Note also that \ts{FeynCalc} already includes different schemes to handle $\gamma_5$, and is expected to improve the treatment of $\gamma_5$ in dimensional regularization in future versions.}

\n Having clarified the functions in Table \ref{tab:FCfunc}, we must consider a new table, Table \ref{tab:FCreno}, which contains some useful variables concerning renormalization.
Two notes should be added. First, in the \t{CT}\textit{process} variable, \textit{process} corresponds to the names of the incoming and the outgoing particles joined together 
\begin{table}[!h]%
\begin{normalsize}
\normalsize
\begin{center}
\begin{tabular}
{@{\hspace{3mm}}>{\raggedright\arraybackslash}p{2.5cm}>{\raggedright\arraybackslash}p{12.0cm}@{\hspace{3mm}}}
\hlinewd{1.1pt}
Function & Meaning \\
\hline\\[-1.5mm]
\t{CT}\textit{process} & expression containing the total counterterm for the \textit{process} at stake \\[2.5mm]
\t{PreResReno} & sum of the total divergent part and \t{CT}\textit{process} \\[2.5mm]
\t{CTfinlist} & list with all the counterterms computed so far in $\overline{\text{MS}}$\\[2.5mm]
\t{ResReno} & the same as \t{PreResReno}, but using counterterms previously stored in CTfinlist\\[6.5mm]
\t{PosResReno} & the same as \t{ResReno}, but using also the counterterms calculated for the process at stake\\[2.5mm]
\hlinewd{1.1pt}
\end{tabular}
\end{center}
\vspace{-5mm}
\end{normalsize}
\caption{Useful variables concerning renormalization. See text for details.}
\label{tab:FCreno}
\end{table}
\normalsize
(for example, in the SM, for the process $h \to Z\gamma$, \t{CT}\textit{process} is \t{CTHZA}). Second, \t{PosResReno} should always be zero, since in the $\overline{\text{MS}}$ scheme the divergents parts are exactly absorbed by the counterterms.

\n Some final comments on the \ts{FeynCalc} notebook.
First, the indices of the particles are described in Table \ref{tab:FCconv},
\begin{table}[!h]%
\begin{normalsize}
\normalsize
\begin{center}
\begin{tabular}
{@{\hspace{3mm}}
>{}m{1.7cm}
>{\raggedright\arraybackslash}m{1.5cm}
>{\raggedright\arraybackslash}m{1.3cm}
>{\raggedright\arraybackslash}m{1.7cm}
>{\raggedright\arraybackslash}m{1.3cm}
>{\raggedright\arraybackslash}m{1.3cm}
>{\raggedright\arraybackslash}m{1.7cm}
>{\raggedright\arraybackslash}m{1.6cm}
@{\hspace{3mm}}}
\hlinewd{1.1pt}
& \small order in \t{Control.m} & \small \ts{QGRAF} index & \small Lorentz index in \ts{FeynCalc} & \small Lorentz index in \LaTeX \,& \small Color index in \LaTeX \,& \small Momentum index in \ts{FeynCalc} & \small Momentum index in \LaTeX \,\\
\hline\\[-2.5mm]
\multirow{2}{2.5cm}{\\[-3.5mm]Incoming particles} & 
\multirow{2}{2.5cm}{1\\[1.5mm]2} &
\multirow{2}{2.5cm}{-1\\[1.5mm]-3} & \multirow{2}{2.5cm}{\t{-J1}\\[1.5mm]\t{-J3}} & \multirow{2}{2.5cm}{$\mu$\\[1.5mm]$\rho$} & 
\multirow{2}{2.5cm}{$a$\\[1.5mm]$c$} & 
\multirow{2}{2.5cm}{\t{p1}\\[1.5mm]\t{p2}} & \multirow{2}{2.5cm}{$p_1$\\[1.5mm]$p_2$}\\[8mm]
\multirow{2}{2.5cm}{\\[-3.5mm]Outgoing particles}  &
\multirow{2}{2.5cm}{1\\[1.5mm]2} &
\multirow{2}{2.5cm}{-2\\[1.5mm]-4} & \multirow{2}{2.5cm}{\t{-J2}\\[1.5mm]\t{-J4}} & \multirow{2}{2.5cm}{$\nu$\\[1.5mm]$\sigma$} &
\multirow{2}{2.5cm}{$b$\\[1.5mm]$d$} &
\multirow{2}{2.5cm}{\t{q1}\\[1.5mm]\t{q2}} & \multirow{2}{2.5cm}{$q_1$\\[1.5mm]$q_2$}\\[8mm]
\hlinewd{1.1pt}
\end{tabular}
\end{center}
\vspace{-5mm}
\end{normalsize}
\caption{Particle indices for the \ts{FeynCalc} notebook. }
\label{tab:FCconv}
\end{table}
\normalsize
and momentum conservation can be applied through the replacement rule \t{MomCons}.\fn{Whenever there is one incoming particle and two outgoing particles, \t{MomCons} replaces \t{p1} by the remaining momenta; in all the other cases, MomCons replaces \t{q1} by the remaining momenta.}
Second, the \t{Helper.m} file contains, among other definitions, both the \t{factor} (in case it was defined in \t{Control.m}) as well as the \ts{FeynCalc} identities (in case they were defined as \t{M\$FCeqs} in the \ts{FeynMaster} model).
Third, even if \t{Comp} is set to \t{False} in \t{Control.m}, setting \t{SumLogic} and \t{MoCoLogic} to \t{True} has consequences for the \ts{FeynCalc} notebook:
when the notebook is run with \t{compNwrite} set to \t{True}, the total expressions will be calculated and momentum conservation will be applied, respectively.
Finally, the replacement rule \t{FCsimp} contains the simplifications for \ts{FeynCalc} (in case they were defined as \t{M\$FCsimp} in the \ts{FeynMaster} model) and is applied in the calculation of \t{res}, \t{resD}, \t{restot} and \t{resDtot}.

\section{Examples}
\label{sec:Examples}

\subsection{Creation and complete automatic renormalization of a toy model}
\label{sec:FullReno}

\n Here we exemplify how to create a model, on the one hand, and how to completely renormalize it, on the other. The model will be very simple: QED with an extra fermion. We first show how to create such a toy model.

\n Probably the simplest way to create any model whatsoever is to copy and modify an already existing model. Given the similarity between our toy model and QED, we go to the directory with models for \FM and
duplicate the directory \t{QED}, after which we name the duplicate \t{QED2}. We get inside \t{QED2} and change the name of the model file from \t{QED.fr} to \t{QED2.fr}. We then open \t{QED2.fr} and edit the model in three steps:
first, we modify the parameter list to:
\vs{2mm}
\begin{small}
\begin{addmargin}[8mm]{0mm}
\begin{verbatim}
(***** Parameter list *****)
M$Parameters = {
  m1 == {TeXName -> "m_1", Renormalization -> {m1 -> m1 + dm1}},
  m2 == {TeXName -> "m_2", Renormalization -> {m2 -> m2 + dm2}},
  ee == {TeXName -> "e", Renormalization -> {ee -> ee + de ee}},
  xiA == {TeXName -> "\\xi_A"},
  de == {Counterterm -> True, TeXName -> "\\delta e"},
  dZ3 == {Counterterm -> True, TeXName -> "\\delta Z_3"},
  dZ1L == {Counterterm -> True, TeXName -> "\\delta Z_1^L"},
  dZ1R == {Counterterm -> True, TeXName -> "\\delta Z_1^R"},
  dZ2L == {Counterterm -> True, TeXName -> "\\delta Z_2^L"},
  dZ2R == {Counterterm -> True, TeXName -> "\\delta Z_2^R"},  
  dm1 == {Counterterm -> True, TeXName -> "\\delta m_1"},
  dm2 == {Counterterm -> True, TeXName -> "\\delta m_2"}};
\end{verbatim}
\end{addmargin}
\end{small}
\vs{3mm}
Then, in the particle classes list, we slightly modify what we had, and we add a second fermion:\fn{The fermions are defined both in terms of Weyl spinors (the \t{W} variables) and Dirac spinors (the \t{F} variables). It is certainly true that, in models with no parity violation (like the present one), there is no need to define the fermions in terms of Weyl spinors. Nevertheless, we consider them for illustrative purposes.}
\vs{5mm}
\begin{small}
\begin{addmargin}[8mm]{0mm}
\begin{verbatim}
(***** Particle classes list *****)
M$ClassesDescription = {
  W[1] == {
		ClassName -> psi1L,
		SelfConjugate -> False,
		QuantumNumbers -> {Q-> Q},
		Renormalization -> {psi1L -> psi1L + 1/2 dZ1L psi1L},
		Chirality -> Left},
  W[2] == {
		ClassName -> chi1R,
		SelfConjugate -> False,
		QuantumNumbers -> {Q-> Q},
		Renormalization -> {chi1R -> chi1R + 1/2 dZ1R chi1R},
		Chirality -> Right},
  W[3] == {
		ClassName -> psi2L,
		SelfConjugate -> False,
		QuantumNumbers -> {Q-> Q},
		Renormalization -> {psi2L -> psi2L + 1/2 dZ2L psi2L},
		Chirality -> Left},
  W[4] == {
		ClassName -> chi2R,
		SelfConjugate -> False,
		QuantumNumbers -> {Q-> Q},
		Renormalization -> {chi2R -> chi2R + 1/2 dZ2R chi2R},
		Chirality -> Right},		
  F[1] == {
		ClassName -> f1,
		TeXName -> "f_1",
		TeXAntiName -> "\\bar{f_1}",
		SelfConjugate -> False,
		QuantumNumbers -> {Q-> Q},
		Mass -> m1,
		WeylComponents -> {psi1L, chi1R}},
  F[2] == {
		ClassName -> f2,
		TeXName -> "f_2",
		TeXAntiName -> "\\bar{f_2}",
		SelfConjugate -> False,
		QuantumNumbers -> {Q-> Q},
		Mass -> m2,
		WeylComponents -> {psi2L, chi2R}},			
  V[1] == {
		ClassName -> A,
		TeXName -> "\\gamma",
		Renormalization -> {A[mu_] -> A[mu] + 1/2 dZ3 A[mu]},
		Mass -> 0,
		SelfConjugate -> True}};
\end{verbatim}
\end{addmargin}
\end{small}
\vs{3mm}
We modify the Lagrangian to include a second fermion:
\vs{3mm}
\begin{small}
\begin{addmargin}[8mm]{0mm}
\begin{verbatim}
LGauge := -1/4 FS[A, \[Mu], \[Nu]] FS[A, \[Mu], \[Nu]]
LFermions := I psi1Lbar.sibar[mu].del[psi1L, mu] + I chi1Rbar.si[mu].del[chi1R, mu] \
		   - m1 (psi1Lbar.chi1R + chi1Rbar.psi1L) \
		   + ee psi1Lbar.sibar[mu].psi1L A[mu] + ee chi1Rbar.si[mu].chi1R A[mu] \
		   + I psi2Lbar.sibar[mu].del[psi2L, mu] + I chi2Rbar.si[mu].del[chi2R, mu] \
		   - m2 (psi2Lbar.chi2R + chi2Rbar.psi2L) \
		   + ee psi2Lbar.sibar[mu].psi2L A[mu] + ee chi2Rbar.si[mu].chi2R A[mu]
LGF := -1/2/xiA del[A[mu], mu] del[A[nu], nu]			 
\end{verbatim}
\end{addmargin}
\end{small}
\vs{3mm}
This completes the model. Now, we want to proceed to its complete automatic renormalization --- that is, to the determination of the analytical expressions for the complete set of counterterms (in the $\overline{\text{MS}}$ scheme).
To do so, we open \t{Control.m}; we start by setting \t{model:\,\,QED2}. Then, we must choose a sequence of processes such that all the counterterms are computed. To do so, note that the total set of counterterms is:
\be
\delta Z_3, \quad \delta Z^{L}_1, \, \delta Z^{R}_1, \, \delta m_1, \quad \delta Z^{L}_2, \, \delta Z^{R}_2, \, \delta m_2, \quad \delta e.
\ee
However, from the renormalization of QED, we know that the first one, $\delta Z_3$, can be determined by the vacuum polarization of the photon; the following three, $\delta Z^{L}_1, \, \delta Z^{R}_1, \, \delta m_1$, can be determined by the self-energy of $f_1$; by the same token, $\delta Z^{L}_2, \, \delta Z^{R}_2, \, \delta m_2$ can be determined by the self-energy of $f_2$; finally, $\delta e$ can be determined by one of the vertices (either $f_1 \bar{f}_1 \gamma$ or $f_2 \bar{f}_2 \gamma$) at 1 loop. Therefore, we write:
\vs{2mm}
\begin{small}
\begin{addmargin}[10mm]{0mm}
\begin{Verbatim}[commandchars=\\\{\}]
inparticles: A
outparticles: A
loops: 1

inparticles: f1
outparticles: f1
loops: 1
options: onepi

inparticles: f2
outparticles: f2
loops: 1
options: onepi

inparticles: A
outparticles: f1,f1bar
loops: 1
options: onepi
\end{Verbatim}
\end{addmargin}
\end{small}
\vs{1mm}
Finally, concerning the logical variables of \t{Control.m}, we set them all to \t{True}, except \t{LoSpinors}, which we set to \t{False}. This being done, everything is set. We then go to the \t{FeynMaster} folder and run batch the file \t{RUN-\ts{FeynMaster}}. In total, 10 PDF files are automatically and subsequentially generated and opened: one for the tree-level Feynman rules, another one for the counterterms Feynman rules, and two files per process --- one with the Feynman diagrams, another with the respective expressions. In the last file for the expressions, we read ``\textit{This completes the renormalization of the model}'', and the list of the full set of counterterms is presented.

\subsection{$h \to \gamma\gamma$ in the Standard Model}
\label{sec:HAA}

\n In this example, we use the $h \to \gamma\gamma$ in the SM as an illustration of several features of \ts{FeynMaster}.
We use the SM model file that comes with \FMM. As for \t{Control.m}, we set it as:%
\fn{\label{note:TrueFalse} We are setting \t{FRinterLogic} to \t{True}, which only needs to be done in case it was not yet done before. Actually, generating all Feynman rules for both the tree-level interactions and the counterterms in the SM may take around 5 minutes. Therefore, if we have already performed that operation, we can save time by setting \t{FRinterLogic} and \t{RenoLogic} to \t{False}.}
\vs{2mm}
\begin{small}
\begin{addmargin}[10mm]{0mm}
\begin{Verbatim}[commandchars=\\\{\}]
model: SM

inparticles: H
outparticles: A,A
loops: 1

FRinterLogic: T
RenoLogic: T
Draw: T
Comp: F
FinLogic: F
DivLogic: F
SumLogic: T
MoCoLogic: F
LoSpinors: F
\end{Verbatim}
\end{addmargin}
\end{small}
%
We then run the batch file \t{RUN-FeynMaster}. In total, 3 PDF files will automatically be generated and opened: one for the tree-level Feynman rules, another one for the counterterms Feynman rules and a third one for the Feynman diagrams. We go to the directory \t{SM/Processes/1-HAA} (meanwhile generated inside the directory for the \FM output) and open the \ts{FeynCalc} notebook \t{Notebook.nb}. We then run the \t{Notebook.nb}, after which we are ready to test some relevant features.

\subsubsection{Notebook access to Feynman rules}

\n First, we want to gain some intuition on how the notebook has access to the SM Feynman rules and to the amplitudes of the $h \to \gamma\gamma$ decay. We write
\vs{1mm}
\begin{small}
\begin{addmargin}[8mm]{0mm}
\begin{Verbatim}[commandchars=\\\{\}]
\textcolor{mygray}{In[14]:=} amp1
\end{Verbatim}
\end{addmargin}
\end{small}
\vs{-2mm}
which should yield the expression:
\bd
\dfrac{2 \, e^3 \, {m_W} \, g^{-J2-J4}}{s_w \left(k_1^2 - {{m_W}}^2\right) \left((p_1-k_1)^2 - {{m_W}}^2\right)} \,  \,  - \dfrac{2 \, D \, e^3 \, {m_W} \, g^{-J2-J4}}{s_w \left(k_1^2 - {{m_W}}^2\right) \left((p_1-k_1)^2 - {{m_W}}^2\right)}.
\ed
This is the amplitude for the first Feynman diagram, where $D$ represents the dimension. Now, where does the notebook get this information from? To answer the question, we open \t{Amplitudes.m} inside \t{1-HAA}. If we check the first line, we realize that \t{amp1} is essentially a product of Feynman rules such as \t{propWP[...]} and \t{vrtxAAWPWPbar[...]}.\fn{Amplitudes 13 to 22 have an extra factor 3, which corresponds to the color number. This factor is automatically added by \FM for diagrams with closed loops of quarks.} These rules are defined in the auxiliary files inside the directory \t{SM/FeynmanRules} (inside the directory for the \FM output). Although they have been automatically generated, they can always be edited for particular purposes.

\subsubsection{Finiteness and gauge invariance}
\label{sec:FT}

\n Next, we use some of the features described in section \ref{sec:FC} to test two important properties of $h \to \gamma\gamma$: finiteness and gauge invariance. We start with the former; by writing
\vs{1mm}
\begin{small}
\begin{addmargin}[8mm]{0mm}
\begin{Verbatim}[commandchars=\\\{\}]
\textcolor{mygray}{In[15]:=} resD
\end{Verbatim}
\end{addmargin}
\end{small}
\vs{-1mm}
we obtain the list with all the expressions for the divergents parts. It is a non-trivial list: although some of its elements are zero, most of them are not. However, when we sum the whole list, we find:
\vs{1mm}
\begin{small}
\begin{addmargin}[8mm]{0mm}
\begin{Verbatim}[commandchars=\\\{\}]
\textcolor{mygray}{In[16]:=} resDtot
\end{Verbatim}
\vs{-3mm}
\begin{Verbatim}[commandchars=\\\{\}]
\textcolor{mygray}{Out[16]=} 0
\end{Verbatim}
\end{addmargin}
\end{small}
\vs{-2mm}
so that the process as a whole is finite, as expected for this decay mode.

\n Let us now check gauge invariance. First of all, note that the total amplitude $M$ for $h \to \gamma\gamma$ can be written as
\be
M = \epsilon_1^{\nu} \epsilon_2^{\sigma} \, M_{\nu\sigma},
\ee
where we are just factoring out the polarization vectors $\epsilon_1^{\nu}$ and $\epsilon_2^{\sigma}$ of the two photons. Then, it is easy to show that gauge invariance forces $M^{\nu\sigma}$ to have the form
\be
M^{\nu\sigma} = ( g^{\nu\sigma} q_1 . q_2 - q_1^{\sigma}  q_2^{\nu}) F,
\label{eq:invgaugefinal}
\ee
where $q_1$ and $q_2$ are the 4-momenta of the two photons, and $F$ is a scalar function of the momenta and the masses. In other words, it is a consequence of gauge invariance that, in the total process, the coefficient of $g^{\nu\sigma} q_1 . q_2$ must be exactly opposite to that of $q_1^{\sigma}  q_2^{\nu}$. To test this, we define some replacement rules:
\begin{small}
\vs{1mm}
\begin{addmargin}[8mm]{0mm}
\begin{Verbatim}[commandchars=\\\{\}]
\hs{12mm} \textcolor{uglyblue}{(* momentum conservation in scalar products and four-vectors *)}
\end{Verbatim}
\vs{-3mm}
\begin{Verbatim}[commandchars=\\\{\}]
\textcolor{mygray}{In[17]:=} dist = \{SP[p1, x_] -> SP[q1, x] + SP[q2, x], FV[p1, x_] -> FV[q1, x] + FV[q2, x]\};
\end{Verbatim}
\vs{-3mm}
\begin{Verbatim}[commandchars=\\\{\}]
\hs{12mm} \textcolor{uglyblue}{(* external particles on-shell *)}
\end{Verbatim}
\vs{-3mm}
\begin{Verbatim}[commandchars=\\\{\}]
\textcolor{mygray}{In[18]:=} onshell = \{SP[q1, q1] -> 0, SP[q2, q2] -> 0, SP[p1, p1] -> mH^2\};
\end{Verbatim}
\vs{-3mm}
\begin{Verbatim}[commandchars=\\\{\}]
\hs{12mm} \textcolor{uglyblue}{(* kinematics *)}
\end{Verbatim}
\vs{-3mm}
\begin{Verbatim}[commandchars=\\\{\}]
\textcolor{mygray}{In[19:=} kin = \{SP[q1, q2] -> MH^2/2, SP[p1, q1] -> MH^2/2, SP[p1, q2] -> MH^2/2\};
\end{Verbatim}
\vs{-3mm}
\begin{Verbatim}[commandchars=\\\{\}]
\hs{12mm} \textcolor{uglyblue}{(* transversality of the external photons polarizations *)}
\end{Verbatim}
\vs{-3mm}
\begin{Verbatim}[commandchars=\\\{\}]
\textcolor{mygray}{In[20]:=} transv = \{FV[q1, -J2] -> 0, FV[q2, -J4] -> 0\};
\end{Verbatim}
\end{addmargin}
\vs{1mm}
\end{small}
which we use to define a new \t{res} list:
\vs{3mm}
\begin{small}
\begin{addmargin}[8mm]{0mm}
\begin{Verbatim}[commandchars=\\\{\}]
\textcolor{mygray}{In[21]:=} resnew = (res /. dist /. onshell /. kin /. transv) // Simplify;
\end{Verbatim}
\end{addmargin}
\end{small}
\vs{1mm}
Finally, we write the coefficients of $g^{\nu\sigma} q_1 . q_2$ and $q_1^{\sigma}  q_2^{\nu}$ as
\vs{3mm}
\begin{small}
\begin{addmargin}[8mm]{0mm}
\begin{Verbatim}[commandchars=\\\{\}]
\textcolor{mygray}{In[22]:=} resnewT = (Coefficient[resnew, MT[-J2, -J4]] // MyPaVeReduce) 
\hs{12mm}  /(MH^2/2) // Simplify // FCE;
\end{Verbatim}
\vs{-3mm}
\begin{Verbatim}[commandchars=\\\{\}]
\textcolor{mygray}{In[23]:=} resnewL = (Coefficient[resnew, FV[q1, -J4]*FV[q2,-J2]] // MyPaVeReduce)
\hs{12mm}  // Simplify // FCE;
\end{Verbatim}
\end{addmargin}
\end{small}
\vs{1mm}
respectively, to conclude that
\vs{3mm}
\begin{small}
\begin{addmargin}[8mm]{0mm}
\begin{Verbatim}[commandchars=\\\{\}]
\textcolor{mygray}{In[24]:=} Total[resnewT] + Total[resnewL] // Simplify
\end{Verbatim}
\vs{-3mm}
\begin{Verbatim}[commandchars=\\\{\}]
\textcolor{mygray}{Out[24]:=} 0
\end{Verbatim}
\end{addmargin}
\end{small}
in accordance with gauge invariance. For what follows, it is convenient to save the expressions for the total transverse and longitudinal part. We write
\vs{3mm}
\begin{small}
\begin{addmargin}[8mm]{0mm}
\begin{Verbatim}[commandchars=\\\{\}]
\textcolor{mygray}{In[25]:=} FT = Total[resnewT] // Simplify;
\end{Verbatim}
\vs{-3mm}
\begin{Verbatim}[commandchars=\\\{\}]
\textcolor{mygray}{In[26]:=} FL = Total[resnewL] // Simplify;
\end{Verbatim}
\end{addmargin}
\end{small}

\subsubsection{\t{MyTeXForm}}

\n We now illustrate how to use \t{MyTeXForm} inside the \ts{FeynCalc} notebook. Suppose we want to write the sum of final results for the diagrams with quartic vertices (diagrams 1 to 6)  in a \LaTeX \, document. We define the variable \t{toprint1} as
\vs{3mm}
\begin{small}
\begin{addmargin}[8mm]{0mm}
\begin{Verbatim}[commandchars=\\\{\}]
\textcolor{mygray}{In[27]:=} toprint1 = Sum[res[[i]], {\{i, 1, 6\}}] // Simplify
\end{Verbatim}
\end{addmargin}
\end{small}
after which we write
\vs{3mm}
\begin{small}
\begin{addmargin}[8mm]{0mm}
\begin{Verbatim}[commandchars=\\\{\}]
\textcolor{mygray}{In[28]:=} toprint1 // MyTeXForm
\end{Verbatim}
\end{addmargin}
\end{small}
If we now copy the outcome as plain text and paste it in a \LaTeX \, document like the present one, we get:
\bd
- \left( e^3 \,  \left(  \left( m_h^2 + 6 \, m_W^2 \right)  \, B_0\left(p_1^2, m_W^2, m_W^2\right) + m_W^2 \,  \left( -4 + B_0\left(q_1^2, m_W^2, m_W^2\right) + B_0\left(q_2^2, m_W^2, m_W^2\right) \right)  \right)  \, g^{\nu \sigma} \right) / \left( 16 \, {m_W} \, \pi^2 \, {s_w} \right)
\ed
Note that we did not need to break the line manually in the \LaTeX \, equation. This is because we are using the \t{breqn} package, which automatically breaks lines in equations.\fn{For documentation, cf. \url{https://www.ctan.org/pkg/breqn} . Recall that the line breaking does not work when the point where the line is to be broken is involved in three or more parentheses.}

\subsubsection{\t{\ts{Fortran}} interface}

\n We mentioned in the Introduction that \FM includes a numerical interface with \t{\ts{Fortran}}.
We now show how it works in the context of $h \to \gamma\gamma$.
Suppose we want to plot the decay width as a function of the Higgs mass; we could start by computing the total $h \to \gamma\gamma$ decay width:
\vs{3mm}
\begin{small}
\begin{addmargin}[8mm]{0mm}
\begin{Verbatim}[commandchars=\\\{\}]
\textcolor{mygray}{In[27]:=} X0 = restot // DecayWidth
\end{Verbatim}
\end{addmargin}
\end{small}
\vs{-0mm}
However, although this works, it takes a long time and produces large expressions. It is simpler to exploit the generic form of eq.~\ref{eq:invgaugefinal} and use a form factor; that is,
\vs{3mm}
\begin{small}
\begin{addmargin}[8mm]{0mm}
\begin{Verbatim}[commandchars=\\\{\}]
\textcolor{mygray}{In[27]:=} X0 = FTaux (MT[-J2, -J4] MH^2/2 -  FV[q1, -J4] FV[q2, -J2]) // DecayWidth 
\end{Verbatim}
\end{addmargin}
\end{small}
where \t{FTaux} is an auxiliary variable that, in the end, must be replaced by the absolute value of the total transverse part \t{FT}, defined above. Then, we write
\vs{3mm}
\begin{small}
\begin{addmargin}[8mm]{0mm}
\begin{Verbatim}[commandchars=\\\{\}]
\textcolor{mygray}{In[28]:=} (X0 /. FTaux -> Abs[FT] // Simplify) // FCtoFT
\end{Verbatim}
\end{addmargin}
\end{small}
where we performed the referred replacement.%
\fn{Alternatively, following the instructions in note \ref{note:FCtoFT}, we could also have written \t{FCtoFT[X0, \{FTaux -> Abs[FT]\}]}.}
As explained in section \ref{sec:FC}, the command \t{FCtoFT} generates five files: \t{MainFT.F}, \t{Myexp.F}, \t{MyVariables.h}, \t{MyParameters.h} and \t{Makefile-template}. We open \t{MainFT.F} and, immediately after the comments \textit{Write now the rest of the program},
we write%
\footnote{The parameters loaded from the file \t{MyParameter.h} cannot be changed inside the \ts{Fortran} program (\t{MainFT.F}). Hence, since we define the parameter \t{MH} as the Higgs boson mass, we name \t{xMH} the variable we are using to make the plot; in doing so, we must be careful to replace \t{MH} for \t{xMH} in the arguments of \t{Myexp} inside the loop.}
\vs{1mm}
\begin{small}
\begin{addmargin}[8mm]{0mm}
\begin{Verbatim}[commandchars=\\\{\}]
xMH=38d0
do i=1,162
xMH=xMH+1d0
write(50,98)xMH,Myexp(..., xMH, ...)
enddo
\end{Verbatim}
\end{addmargin}
\end{small}
%
where \t{50} and \t{98} represent the output file and the impression format, respectively. We are varying the Higgs mass from 38 GeV to 200 GeV in steps of 1 GeV. The result is presented in Fig.~{\ref{fig:HAA}.
\begin{figure}[htb]
\centering
\includegraphics[width=0.5\textwidth]{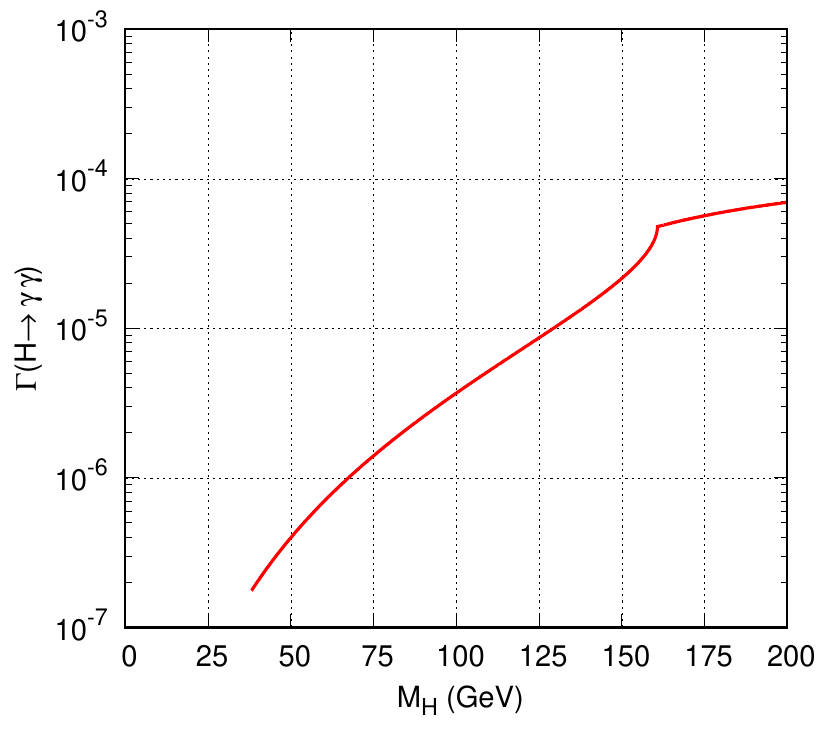}
\vs{-5mm}
\caption{Width of the process $h \to \gamma \gamma$ as a function of the Higgs boson mass.}
\label{fig:HAA}
\end{figure}	

\vs{-3mm}
\subsubsection{Edition of Feynman diagrams}

\n Finally, we briefly explain how to edit the Feynman diagrams. Recall that they were written in a \LaTeX \, file inside \t{SM/1-HAA/TeXs-drawing} folder. We open the file \t{diagrams.tex} and consider the first diagram; the original code produces the original diagram:

\vs{3mm}
\hspace{7mm} 
\begin{minipage}[h]{.80\textwidth}
\begin{small}
\begin{verbatim}
(...)
\fmflabel{$\gamma$}{...} 
\fmflabel{$\gamma$}{...} 
\fmf{dashes,tension=3}{...} 
\fmf{photon,tension=3}{...} 
\fmf{photon,tension=3}{...} 
\fmf{photon,label=$W^{+}$,right=1}{...}
\fmf{phantom_arrow,tension=0,right=1}{...}
\fmf{photon,label=$W^{+}$,right=1}{...}
\fmf{phantom_arrow,tension=0,right=1}{...}
(...)
\end{verbatim}
\end{small}
\end{minipage}
\hspace{-45mm} 
\begin{minipage}[h]{.40\textwidth}
\begin{picture}(0,80)
\begin{fmffile}{1} 
\begin{fmfgraph*}(100,56) 
\fmfset{arrow_len}{3mm} 
\fmfset{arrow_ang}{20} 
\fmfleft{nJ1} 
\fmflabel{$h$}{nJ1} 
\fmfright{nJ2,nJ4} 
\fmflabel{$\gamma$}{nJ2} 
\fmflabel{$\gamma$}{nJ4} 
\fmf{dashes,tension=3}{nJ1,nJ1J3J2} 
\fmf{photon,tension=3}{nJ2,nJ2nJ4J1J4} 
\fmf{photon,tension=3}{nJ4,nJ2nJ4J1J4} 
\fmf{photon,label=$W^+$,right=1}{nJ1J3J2,nJ2nJ4J1J4} 
\fmf{phantom_arrow,tension=0,right=1}{nJ1J3J2,nJ2nJ4J1J4} 
\fmf{photon,label=$W^+$,right=1}{nJ2nJ4J1J4,nJ1J3J2} 
\fmf{phantom_arrow,tension=0,right=1}{nJ2nJ4J1J4,nJ1J3J2} 
\end{fmfgraph*} 
\end{fmffile}
\end{picture}
\end{minipage}
\vs{1mm}

\n However, we can modify the code in order to change the aspect of the diagram. In particular, we can change the labels, the tensions and the curvatures.\fn{The tensions represent the strength of the lines: the larger the tension, the tighter the line will be. The default tension is 1. The curvature is represented by the variable \t{right}. Note that tensions, labels and curvatures are just a few examples of variables that can be changed to generate a different diagram. For more informations, please consult the \t{feynmf} manual.} For example:

\vs{3mm}
\hspace{7mm} 
\begin{minipage}[h]{.80\textwidth}
\begin{small}
\begin{verbatim}
(...)
\fmflabel{$\gamma_1$}{...} 
\fmflabel{$\gamma_2$}{...} 
\fmf{dashes,tension=1}{...} 
\fmf{photon,tension=3}{...} 
\fmf{photon,tension=3}{...} 
\fmf{photon,label=$W^{-}$,right=1}{...} 
\fmf{phantom_arrow,tension=0,right=1}{...} 
\fmf{photon,label=$W^{-}$,right=1}{...}
\fmf{phantom_arrow,tension=0,right=1}{...} 
(...)
\end{verbatim}
\end{small}
\end{minipage}
\hspace{-45mm} 
\begin{minipage}[h]{.40\textwidth}
\begin{picture}(0,80)
\begin{fmffile}{2} 
\begin{fmfgraph*}(100,56) 
\fmfset{arrow_len}{3mm} 
\fmfset{arrow_ang}{20} 
\fmfleft{nJ1} 
\fmflabel{$h$}{nJ1} 
\fmfright{nJ2,nJ4} 
\fmflabel{$\gamma_1$}{nJ2} 
\fmflabel{$\gamma_2$}{nJ4} 
\fmf{dashes,tension=1}{nJ1,nJ1J3J2} 
\fmf{photon,tension=3}{nJ2,nJ2nJ4J1J4} 
\fmf{photon,tension=3}{nJ4,nJ2nJ4J1J4} 
\fmf{photon,label=$W^{+}$,right=1}{nJ1J3J2,nJ2nJ4J1J4} 
\fmf{phantom_arrow,tension=0,right=1}{nJ1J3J2,nJ2nJ4J1J4} 
\fmf{photon,label=$W^+$,right=1}{nJ2nJ4J1J4,nJ1J3J2} 
\fmf{phantom_arrow,tension=0,right=1}{nJ2nJ4J1J4,nJ1J3J2} 
\end{fmfgraph*} 
\end{fmffile}
\end{picture}
\end{minipage}
\vs{8mm}

\vs{-5mm}
\subsection{QED Ward identity}
\label{sec:WI}

\n In the previous example, we showed how to use \ts{FeynMaster} to manipulate the results of a single process. Here, we illustrate how it can also be used to combine information of several processes. For that purpose, we consider a simple task: prove the QED Ward identity.

\n It is easy to show that the Ward identity at one-loop order in QED can be written as:
\be
p_1^{\nu} \, \Gamma_{\nu}(p_1,p_2,p_3) = e \left( \vphantom{\dfrac{A^B}{A^B}} \Sigma(p_2) - \Sigma(p_3) \right),
\label{eq:provanda}
\ee
where
\be
\Gamma_{\nu}(p_1,p_2,p_3) =
\hs{8mm}
\begin{minipage}{0.35\textwidth}
\begin{fmffile}{QEDB11}
\begin{fmfgraph*}(100,100) 
\fmfset{arrow_len}{3mm} 
\fmfset{arrow_ang}{20} 
\fmfleft{nJ1}  
\fmfright{nJ2,nJ4} 
\fmf{photon,label=$p_1$,tension=4}{nJ1,J2J3nJ1} 
\fmf{fermion,label=$p_2$,label.side=left,tension=4}{nJ2J1J5,nJ2} 
\fmf{fermion,label=$p_3$,label.side=left,tension=4}{nJ4,J4nJ4J6} 
\fmf{fermion,tension=1,label.dist=3thick}{J2J3nJ1,nJ2J1J5} 
\fmf{fermion,tension=1,label.dist=3thick}{J4nJ4J6,J2J3nJ1} 
\fmf{photon,label.side=left,tension=1,label.dist=3thick}{J4nJ4J6,nJ2J1J5} 
\end{fmfgraph*}
\end{fmffile}
\end{minipage}
\hs{-17mm},
\hs{15mm}
\Sigma(p_i) = 
\begin{minipage}{0.35\textwidth}
\begin{fmffile}{QEDB12} 
\begin{fmfgraph*}(100,57) 
\fmfset{arrow_len}{3mm} 
\fmfset{arrow_ang}{20} 
\fmfleft{nJ1} 
\fmfright{nJ2} 
\fmf{fermion,label=$p_i$,tension=3}{nJ1,J2nJ1J3} 
\fmf{fermion,label=$p_i$,label.side=right,tension=3}{nJ2J1J4,nJ2} 
\fmf{fermion,right=1}{J2nJ1J3,nJ2J1J4} 
\fmf{photon,right=1}{nJ2J1J4,J2nJ1J3} 
\end{fmfgraph*} 
\end{fmffile} 
\end{minipage}
\hs{-15mm},
\label{eq:pic}
\ee
and where the momenta $p_1$ and $p_3$ are incoming, while $p_2$ is outgoing.
In order to prove eq. \ref{eq:provanda} with \ts{FeynMaster}, we need to consider the two processes depicted in eq. \ref{eq:pic}: the QED vertex and the fermion self-energy. Hence, we open and edit \t{Control.m} according to:
\vs{2mm}
\begin{small}
\begin{addmargin}[10mm]{0mm}
\begin{Verbatim}[commandchars=\\\{\}]
model: QED

inparticles: A
outparticles: f,fbar
loops: 1
options: onepi

inparticles: f
outparticles: f
loops: 1
options: onepi

FRinterLogic: T
RenoLogic: F
Draw: F
Comp: T
FinLogic: F
DivLogic: F
SumLogic: T
MoCoLogic: F
LoSpinors: F
\end{Verbatim}
\end{addmargin}
\end{small}
%
We then run \ts{FeynMaster}.
After this, we go to the directory \t{QED/Processes/1-Affbar} (meanwhile generated inside the directory for the \FM output), we copy the notebook lying there to a different directory and we rename it \t{Notebook-Global.nb}. This is going to be the notebook where we shall combine the information of both processes. We open it, and delete most of the lines there: in a first phase, we only want to load the general files. So it must look like this:
\vs{1mm}
\begin{small}
\begin{addmargin}[8mm]{0mm}
\begin{Verbatim}[commandchars=\\\{\}]
\textcolor{mygray}{In[1]:=} << FeynCalc`
\textcolor{mygray}{In[2]:=} dirNuc = "(...)"; 
\textcolor{mygray}{In[3]:=} dirFey = "(...)"; 
\textcolor{mygray}{In[4]:=} dirCT = "(...)"; 
\textcolor{mygray}{In[5]:=} Get["Feynman-Rules-Main.m", Path -> \{dirFey\}] 
\textcolor{mygray}{In[6]:=} Get["FunctionsFM.m", Path -> \{dirNuc\}]
\textcolor{mygray}{In[7]:=} compNwrite = False; 
\end{Verbatim}
\end{addmargin}
\end{small}
\vs{-1mm}
Now, we want to load the first process. To do so, we write:
\vs{1mm}
\begin{small}
\begin{addmargin}[8mm]{0mm}
\begin{Verbatim}[commandchars=\\\{\}]
\textcolor{mygray}{In[8]:=} dirHome = "(...)"; 
\textcolor{mygray}{In[9]:=} SetDirectory[dirHome];
\textcolor{mygray}{In[10]:=} << Helper.m;
\textcolor{mygray}{In[11]:=} Get["Definitions.m", Path -> \{dirNuc\}];
\textcolor{mygray}{In[12]:=} Get["Finals.m", Path -> \{dirNuc\}]; 
\end{Verbatim}
\end{addmargin}
\end{small}
\vs{-1mm}
where \t{dirHome} should be set to the directory corresponding to \t{1-Affbar}. Next, we define new variables: \t{X0} as $\Gamma$ of eq. \ref{eq:provanda}, with the above-mentioned momentum definitions, and \t{X1} as the whole left-hand side of eq. \ref{eq:provanda}.
\vs{1mm}
\begin{small}
\begin{addmargin}[8mm]{0mm}
\begin{Verbatim}[commandchars=\\\{\}]
\textcolor{mygray}{In[13]:=} X0 = res[[1]] /. \{p1 -> p2 - p3, q1 -> p2, q2 -> -p3\};
\textcolor{mygray}{In[14]:=} X1 = Contract[X0 FV[p2 - p3, -J1]];
\end{Verbatim}
\end{addmargin}
\end{small}
\vs{-1mm}
We now load the second process:
\vs{1mm}
\begin{small}
\begin{addmargin}[8mm]{0mm}
\begin{Verbatim}[commandchars=\\\{\}]
\textcolor{mygray}{In[15]:=} dirHome = "(...)"; 
\textcolor{mygray}{In[16]:=} SetDirectory[dirHome];
\textcolor{mygray}{In[17]:=} << Helper.m ;
\textcolor{mygray}{In[18]:=} Get["Definitions.m", Path -> \{dirNuc\}];
\textcolor{mygray}{In[19]:=} Get["Finals.m", Path -> \{dirNuc\}]; 
\end{Verbatim}
\end{addmargin}
\end{small}
\vs{-1mm}
where \t{dirHome} should now be set to the directory corresponding to \t{2-ff}. From this, and recalling that the default momentum of a self-energy is \t{p1} (cf. Table \ref{tab:FCconv}), we can obtain the right-hand side of eq. \ref{eq:provanda} by writing:
\vs{1mm}
\begin{small}
\begin{addmargin}[8mm]{0mm}
\begin{Verbatim}[commandchars=\\\{\}]
\textcolor{mygray}{In[20]:=} Y0a = res[[1]] /. {p1 -> p2}; 
\textcolor{mygray}{In[21]:=} Y0b = res[[1]] /. {p1 -> p3};
\textcolor{mygray}{In[22]:=} Y1 = (e (Y0a - Y0b) // DiracSimplify) // Simplify;
\end{Verbatim}
\end{addmargin}
\end{small}
\vs{-1mm}
Finally, we prove the Ward identity by showing that both sides of eq. \ref{eq:provanda} are equal:
\vs{1mm}
\begin{small}
\begin{addmargin}[8mm]{0mm}
\begin{Verbatim}[commandchars=\\\{\}]
\textcolor{mygray}{In[23]:=} WI = (Y1 - X1) // Simplify;
\textcolor{mygray}{In[24]:=} CheckWI = MyPaVeReduce[WI] // Calc
\end{Verbatim}
\end{addmargin}
\end{small}
\vs{-1mm}
which yields 0, thus completing the proof.

\section{Quick first usage}
\label{sec:Summary}

\n For a quick first usage of \ts{FeynMaster}, the user should follow this sequence of steps:
\vs{1.0mm}
\begin{center}
\begin{addmargin}[8mm]{0mm}
1) Make sure you have installed \t{\ts{Python}}, \t{\ts{Mathematica}} and \LaTeX \,, on the one hand, and \ts{FeynRules}, \ts{QGRAF} and \ts{FeynCalc}, on the other;\\[2mm]
2) Download \ts{FeynMaster} in \url{https://porthos.tecnico.ulisboa.pt/FeynMaster/};\\[2mm]
3) Extract the downloaded file and place the resulting folder in a directory at will;\\[2mm]
4) Edit the file \t{RUN-FeynMaster} as explained in section \ref{sec:Instal};\\[2mm]
5) Run \t{RUN-FeynMaster}.
\end{addmargin}
\end{center}
This should generate and open 4 PDF files relative to QED: the Feynman rules for the tree-level interactions, the Feynman rules for the counterterms
interactions, the Feynman diagram for the one-loop vacuum polarization, and a document containing not only the expressions for the vacuum polarization, but also the expression for the associated counterterm in $\overline{\text{MS}}$.



\section*{Acknowledgements}

\n Both authors are very grateful to António P. Lacerda, who kicked off the entire program. We also thank Vladyslav Shtabovenko, Paulo Nogueira and Augusto Barroso for useful discussions concerning \ts{FeynCalc}, \ts{QGRAF} and renormalization, respectively; Maximilian Löschner for bringing the \t{feynmf} package to our attention; Miguel P. Bento and Patrick Blackstone for testing the program; { Darius Jur\v{c}iukonis for} {a careful reading of the manuscript;} João P. Silva for the suggestion of the name `\ts{FeynMaster}', as well as for a careful reading of the manuscript. D.F. is also grateful to Isabel Fonseca for many useful suggestions concerning \t{\ts{Python}} and to Sofia Gomes for a suggestion regarding the printing of the Feynman rules.


\end{document}